# High Energy Emission from Rotation-Powered Pulsars: Outer-gap vs. Slot-gap Models

**Kouichi Hirotani**

ASIAA/National Tsing Hua University, TIARA, P. O. Box 23-141, Taipei, Taiwan

Postal address: TIARA, Department of Physics, National Tsing Hua University, 101, Sec. 2, Kuang Fu Rd., Hsinchu, Taiwan 300

We explore particle accelerator electrodynamics in the magnetosphere of a rapidly rotating neutron star (NS). We address the importance of a self-consistent treatment of pair production, solving the Poisson equation describing the acceleration electric field, the Boltzmann equations for produced electrons and positrons, and the radiative transfer equation simultaneously. It is demonstrated that the accelerator solution is obtained if we only specify the NS spin period, magnetic dipole moment, magnetic inclination angle with respect to the rotation axis, and the NS surface temperature, and that the solution corresponds to a quantitative extension of previous outer-gap models. We apply the scheme to the Crab pulsar and show that the predicted pulse profiles and phase-resolved spectrum are roughly consistent with observations. Applying the same scheme to the slot-gap model, we show that this alternative model predicts too small photon flux to reproduce observations, because the gap trans-field thickness is significantly restricted by its pair-free condition.

## INTRODUCTION

Rotation-powered pulsars are excellent laboratories for testing particle acceleration theories. The major advantage of studying pulsars is that we know that they are powered by rotational energy loss. It is generally accepted that the rotating magnetic field works as a unipolar inductor to exert a large electro-motive force on the spinning neutron star (NS) surface and that the resulting potential drop in the rotating magnetosphere leads to the acceleration of charged particles to ultra-relativistic energies. The challenge is then to investigate where and how they convert this energy into radiation that we observe.

The pulsar magnetosphere can be divided into two zones (Fig. 1). The closed zone is filled with a dense plasma co-rotating with the star, whereas the open zone allows plasmas to flow along the open field lines, escaping to large distances through the light cylinder. Here, the light cylinder is located at distance $\varpi_{\rm LC}=c/\Omega$ from the rotation axis, where $\Omega$ denotes the NS rotational angular frequency and $c$ the speed of light. The last-open field lines, which become parallel to the rotation axis at the light cylinder, form the border of the open magnetic field line bundle. In all the pulsar emission models, particle acceleration and the resulting photon emissions take place within the open zone.

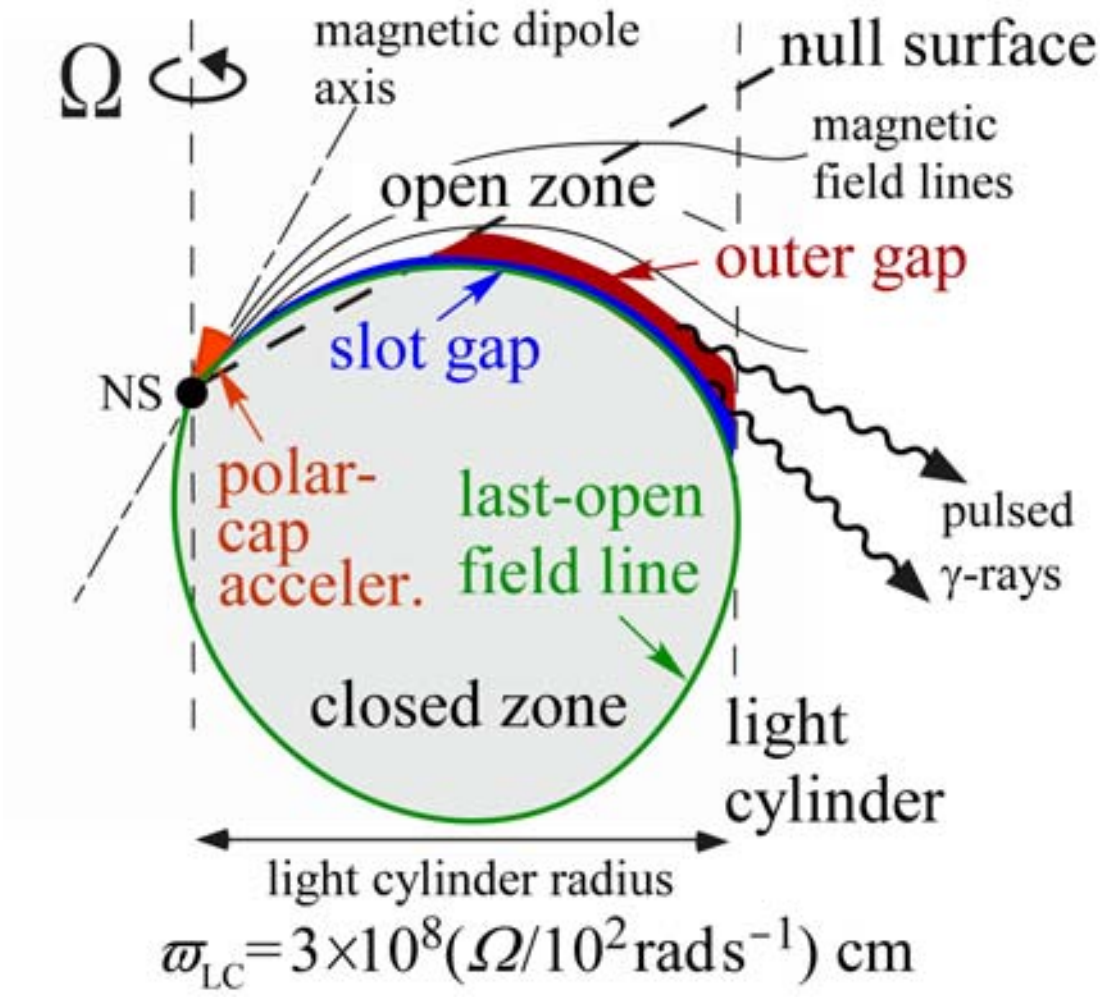

Fig. (1). Schematic figure (side view) of the pulsar magnetosphere. The slot gap extends from the stellar surface to the light cylinder on the last-open field line (in the open zone) with trans-field thickness much less than the outer gap. The outer gap is traditionally assumed to be located between the null charge surface and the light cylinder.

Attempts to model the particle accelerator in a pulsar magnetosphere concentrate on two scenarios: polar-cap (PC) models with emission altitudes within several NS radii above a PC surface (Harding et al. 1978; Daugherty & Harding 1982; Dermer & Sturner 1994; Sturner et al. 1995), and the outer-gap (OG) models with acceleration occurring near the light cylinder (Cheng et al. 1986a, b; Chiang & Romani 1992, 1994, Romani & Yadigaroglu 1995; Romani 1996; Zhang & Cheng 1997; Cheng et al. 2000). Both models predict that electrons and/or positrons are accelerated in a charge depletion region, a potential drop, by the electric field along the magnetic field lines to radiate high-energy gamma-rays via the curvature and inverse-Compton processes.

It is widely accepted from phenomenological studies that *coherent* radio photons are emitted from the PC accelerator. It should be noted that the PC model predicts the so-called 'pencil beam' emission with a small solid angle along the magnetic axis, because the emission takes place only from the lower altitudes (i.e., near the NS surface). Since the radio pulsation generally exhibits a *single* sharp peak in each rotational period, we can reproduce such pulse profiles if the observer's viewing angle is close to the magnetic inclination angle. However, to explain the *incoherent* high energy emissions, which generally exhibit widely separated *double* peaks, one has to invoke a very small magnetic inclination angle and the observer's viewing angle with respect to the rotation axis. Thus, to explain the $\gamma$-ray pulsations detected from all the six brightest pulsars (for Crab, Nolan et al. 1993, Fierro et al. 1998; for Vela, Kanbach et al. 1994; Fierro et al. 1998; for Geminga, Mayer-Hasselwander et al. 1994; Fierro et al. 1994; for PSR B1706-44, Thompson et al. 1996; for B1509-58, Matz et al. 1994; for B1951+32, Ramanamurthy et al. 1995, Kuiper et al. 1998) in terms of the spin-down luminosity, a high-altitude emission, which results in the so-called 'fan beam' emission with a large solid angle (~1 ster), drew attention as an alternative possibility.

The possibility of a high-altitude extension of the PC accelerator was originally introduced by Arons & Scharlemann (ApJ 1979). Developing this electrodynamical model, and incorporating general relativistic effects (Muslimov and Tsygan 1992, hereafter MT92), Muslimov & Harding (2003, hereafter MH03) began discussing high-energy emission from the pulsar slot gap (SG), a narrow region on the last-open field line where the magnetic-field-aligned electric field, $E_{\|}$, is decreasing and accelerating particles cannot create pairs. Then Muslimov & Harding (2004, hereafter MH04) extended this lower-altitude solution (within a few stellar radii) to higher altitudes (near the light cylinder) and found that curvature radiation of the primary particles forms caustic patterns, as suggested in the two-pole caustic model proposed by Dyks & Rudak (2003). Subsequently, Dyks et al. (2004) examined polarization characteristics and found that fast swings of the position angle and minima of polarization degree can be qualitatively reproduced within their two-pole caustic model. This type of emission, SG emission, fills the whole sky and all phases in a light curve. That is, most observers catch emission from the two poles, if their viewing angle, $\zeta_{obs}$, is in a reasonable range (e.g., $45^o<\zeta_{obs}<125^o$ for $\alpha>30^o$, where $\alpha$ denotes the magnetic inclination). More recently, Harding et al. (2008, hereafter HSDF08) investigated the SG emission, adding synchrotron radiation of the gap-accelerated primaries and higher-generation pairs produced at lower-altitude PC region. They applied the model to the Crab pulsar and predicted the pulse profiles and the phase-resolved spectra, which are roughly consistent with the observations from 0.3 keV to 10 GeV.

An alternative way to consider a high-altitude emission is the OG model. To examine the OG quantitatively, Hirotani & Shibata (1999a, b) first solved the set of Maxwell and Boltzmann equations in a pulsar magnetosphere, extending the method originally proposed for a pair-production cascade in a rotating black-hole magnetosphere (Beskin et al. 1992; Hirotani & Okamoto 1998). Before Hirotani & Shibata (1999a), the accelerator position, strength of $E_{\|}$, and even the existence of the OG itself, have been hypothesized; however, they demonstrated, in their one-dimensional analysis of a pair-production cascade along the magnetic field line, that the OG does exist as the solution of the set of the Maxwell and Boltzmann equations. After that, a number of papers appeared in the same context (for one-dimensional solution along the magnetic field lines, see, Hirotani et al. 2003; for two-dimensional solutions, see Takata et al. 2006; Hirotani 2006, 2007a, b; Takata et al. 2008), confirming that the traditional OG models give qualitatively good agreement with these self-consistent solutions. In the present paper, we extend the method presented in Hirotani (2006, hereafter H06), which analyzes the OG electrodynamics in the 2-D poloidal plane and in the 2-D momentum space (neglecting the dependence on toroidal momenta in the distribution functions of particles and photons) into 3-D configuration space and 3-D momentum space. For example, in the present paper, photon propagation and pair production is solved in the full 3-D magnetosphere. We demonstrate that the OG solution is uniquely obtained if we specify the NS rotational period, $P$, magnetic dipole moment, $\mu$, surface temperature, $kT$, and $\alpha$, without introducing any artificial assumptions. Phase-resolved spectra and pulse profiles can be predicted, if we additionally give $\zeta_{obs}$ and the distance to the pulsar, $d$.

In the next section, we present the basic equations. Then in section 3, we demonstrate that the obtained solution quantifies the phenomenological OG models discussed so far. In section 4, we apply the same method to the SG model and point out its difficulties. We finally discuss some implications in section 5.

## 2 BASIC EQUATIONS

*2.1. Background geometry*

Around a rotating NS with angular frequency $\Omega$, mass $M$, and moment of inertia $I$, the background spacetime is given by (Lense & Thirring 1918)

$$ds^2 = g_{tt}dt^2 + 2g_{t\varphi}dtd\varphi + g_{rr}dr^2 + g_{\theta\theta}d\theta^2 + g_{\varphi\varphi}d\varphi^2, \quad (1)$$

where $g_{t\varphi} \equiv ac(r_g/r)\sin^2\theta$ describes the space-time dragging effect, $a \equiv I\Omega/(Mc)$ parameterizes the stellar angular momentum, $g_{tt} \equiv (1-r_g/r)c^2$, $g_{rr} \equiv -(1-r_g/r)^{-1}$, $g_{\theta\theta} \equiv -r^2$, $g_{\varphi\varphi} \equiv -r^2\sin^2\theta$; $r_g \equiv 2GM/c^2$ indicates the Schwarzschild radius. At radius $r$, the inertial frame is dragged at angular frequency

$$\omega \equiv -\frac{g_{t\varphi}}{g_{\varphi\varphi}} = \frac{I}{Mr_*^2}\frac{r_g}{r_*}\left(\frac{r_*}{r}\right)^3 \Omega = \kappa\,\Omega I_{45} r_6^{-3}, \quad (2)$$

where $r_*$ is the stellar radius, $I_{45}=I/(10^{45}\ \mathrm{erg\ cm^2})$, $r_6=r_*/(10^6\ \mathrm{cm})$, and $\kappa \sim 0.15$.

*2.2. Poisson equation for non-corotational potential*

The inhomogeneous part of the Maxwell equations gives

$$\nabla_\mu F^{t\mu} = \frac{1}{\sqrt{-g}}\partial_\mu\left[\frac{\sqrt{-g}}{\rho_w^2} g^{\mu\nu}\left(g_{t\varphi}F_{\varphi\nu} - g_{\varphi\varphi}F_{t\nu}\right)\right] = \frac{4\pi}{c^2}\rho$$

(3)

where $\nabla$ denotes the covariant derivative, the Greek indices run over $t$, $r$, $\theta$, and $\varphi$, $\sqrt{-g} \equiv \sqrt{g_{rr}g_{\theta\theta}\rho_w^2} = cr^2\sin\theta$, and $\rho_w^2 \equiv g_{t\varphi}^2 - g_{tt}g_{\varphi\varphi} = c^2\left(1-r_g/r\right)r^2\sin^2\theta$. If there is an ion emission from the stellar surface, the real charge density $\rho$ is given by $\rho=\rho_e+\rho_{ion}$, where $\rho_e$ denotes the sum of positronic and electronic charge contribution, while $\rho_{ion}$ does the ionic one. The six independent components of the field-strength tensor give the electromagnetic field observed by a distant static observer (Camenzind 1986a, b),

$$E_r = F_{rt},\ E_\theta = F_{\theta t},\ E_\varphi = F_{\varphi t}, \quad (4)$$

$$B^r = \frac{g_{tt}+g_{t\phi}\Omega}{\sqrt{-g}}F_{\theta\varphi},\ B^\theta = \frac{g_{tt}+g_{t\phi}\Omega}{\sqrt{-g}}F_{\varphi r},\ B_\varphi = \frac{-\rho_w^2}{\sqrt{-g}}F_{r\theta}$$

(5)

where $F_{\mu\nu}=A_{\nu,\mu}-A_{\mu,\nu}$ and $A_{\nu,\mu}$ denotes the vector potential $A_\nu$ differentiated with respect to $x^\mu$.

Assuming that the electromagnetic fields are unchanged in the co-rotating frame, we can introduce the non-corotational potential $\Psi$ such that

$$F_{\mu t} + \Omega F_{\mu\phi} = -\partial_\mu\Psi(r,\theta,\varphi-\Omega t), \quad (6)$$

where $\mu=t,r,\theta,\varphi$. If $F_{At}+\Omega A_{A\varphi}=0$ holds for $A=r,\theta$, the angular frequency $\Omega$ of a magnetic field is conserved along the field line. On the NS surface, we impose $F_{\theta t}+\Omega A_{\theta\varphi}=0$ (perfect conductor) to find that the surface is equi-potential, $\partial_\theta\Psi=\partial_t\Psi+\Omega\partial_\varphi\Psi=0$ holds. However, in a particle-acceleration region, $F_{At}+\Omega A_{A\varphi}=0$ deviates from 0 and the magnetic field does not rigidly rotate. The deviation is expressed in terms of $\Psi$, which gives the strength of the acceleration electric field measured by a distant static observer as

$$E_\parallel \equiv \frac{\mathbf{B}\bullet\mathbf{E}}{B} = \frac{B^i}{B}\left(F_{it} + \Omega F_{i\varphi}\right) = \frac{\mathbf{B}}{B}\bullet\left(-\nabla\Psi\right), \quad (7)$$

where the Latin index $i$ runs over spatial coordinates $r,\theta,\varphi$.

Substituting equation (6) into (3), we obtain the Poisson equation for the non-corotational potential,

$$-\frac{c^2}{\sqrt{-g}}\partial_\mu\left(\frac{\sqrt{-g}}{\rho_w^2} g^{\mu\nu} g_{\varphi\varphi}\partial_\nu\Psi\right) = 4\pi\left(\rho - \rho_{GJ}\right), \quad (8)$$

where the Goldreich-Julian (GJ) charge density is defined as (Goldreich & Julian 1969)

$$\rho_{GJ} \equiv \frac{c^2}{4\pi\sqrt{-g}}\partial_\mu\left(\frac{\sqrt{-g}}{\rho_w^2} g^{\mu\nu} g_{\varphi\varphi}\left(\Omega-\omega\right)F_{\varphi\nu}\right). \quad (9)$$

Far from the star, it coincides with the special relativistic expression (Mestel 1971).

In stead of $(r,\theta,\varphi)$, we adopt the magnetic coordinates $(s, \theta_*, \varphi_*)$ such that $s$ denotes the distance along a magnetic field line, $\theta_*$ and $\varphi_*$ represent the magnetic colatitude and azimuth, respectively, of the point where the field line intersects the NS surface. On the NS surface, $\theta$ and $\varphi-\Omega t$ are related with $\theta_*$ and $\varphi_*$ by the following two equations:

$$\cos\theta_* = \sin\alpha\sin\theta\cos(\varphi-\Omega t) + \cos\alpha\cos\theta, \quad (10)$$

$$\cos\theta = -\sin\alpha\sin\theta_*\cos\varphi_* + \cos\alpha\cos\theta_*. \quad (11)$$

For example, if the NS is rotating with period $P$=33 ms and magnetic inclination angle $\alpha=60^\circ$, the angular coordinates $(\theta_*, \varphi_*)$ distribute on the polar cap (PC) surface as depicted in figure 2; $\varphi_*$ is measured counter-clockwise from the $x$ axis, where the rotation axis resides in the $-x$ direction (i.e., leftwards). The last-open field line forms the boundary of the PC region with $\theta_*=\theta_*^{max}(\varphi_*)$, whereas the magnetic axis corresponds to $\theta_*=0$.

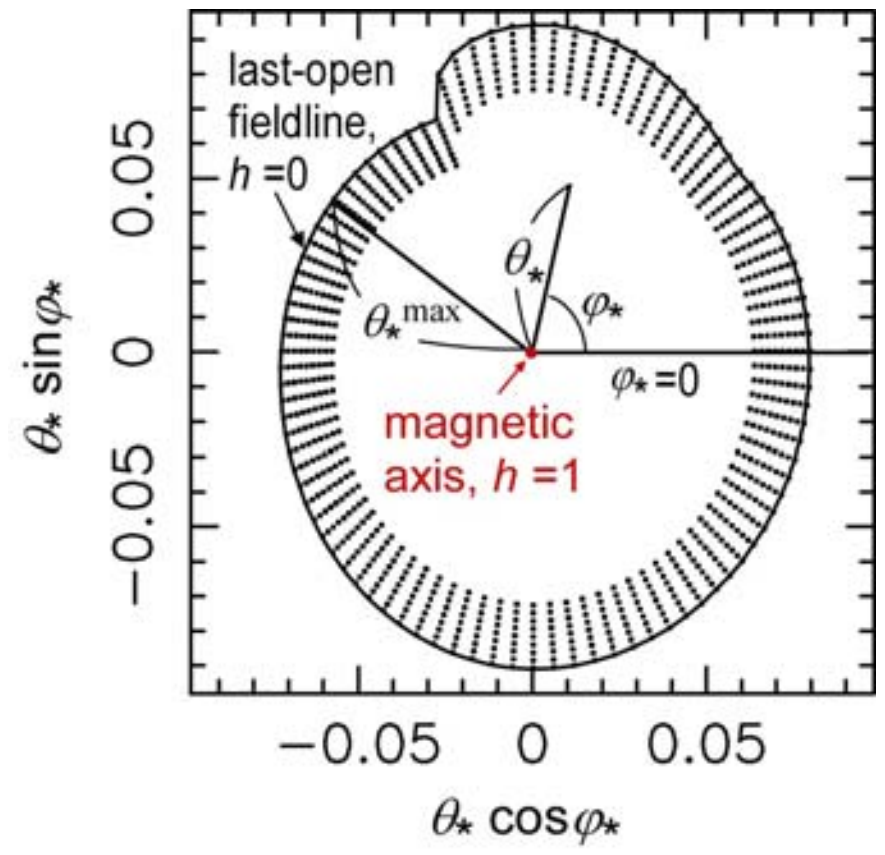

Fig (2). Magnetic coordinates ($\theta_*$, $\varphi_*$) that specify individual magnetic field lines on the polar-cap surface. The dots denote the footpoints of the magnetic field lines along which the simulation is performed; 96 grids were taken in $\varphi_*$ direction and 24 grids in $\theta_*$ direction (even though only 8 points are depicted at each $\varphi_*$).

At distance $s$ from the NS surface, $(r,\theta,\varphi)$ is related with $(s, \theta_*,\varphi_*)$ by

$$r(s,\theta_*,\varphi_*) = r_* + \int_0^s \frac{B^r(s',\theta,\varphi-\Omega t)}{B(s',\theta,\varphi-\Omega t)}ds', \quad (12)$$

$$\theta(s,\theta_*,\varphi_*) = \theta(0,\theta_*,\varphi_*) + \int_0^s \frac{B^\theta(s',\theta,\varphi-\Omega t)}{B(s',\theta,\varphi-\Omega t)}ds', \quad (13)$$

$$\varphi(s,\theta_*,\varphi_*) - \Omega t = \varphi_* + \int_0^s \frac{B^\varphi(s',\theta,\varphi-\Omega t)}{B(s',\theta,\varphi-\Omega t)}ds', \quad (14)$$

where $\theta(0, \theta_*,\varphi_*)$ coincides with $\theta$ in equation (10). Using the coordinates $(s, \theta_*,\varphi_*)$, the Poisson equation for the electrostatic potential $\Psi$ becomes (H06)

$$-\frac{c^2 g_{\varphi\varphi}}{\rho_w^2} g^{i'j'}\partial_{i'}\partial_{j'}\Psi - A^{i'}\partial_{i'}\Psi = 4\pi(\rho - \rho_{GJ}), \quad (15)$$

where the Latin indexes $i'$ and $j'$ runs over $s$, $\theta_*,\varphi_*$; $A^s$, $A^{\theta*}$, and $A^{\varphi*}$ are explicitly defined by equations (A7)-(A9) in H06. The magnetic field expansion effects are described by the trans-field derivative terms in the left-hand side. It follows from equation (7) that the acceleration electric field is given as

$$E_\parallel = -\left(\partial\Psi / \partial s\right)_{\theta_*\varphi_*}. \quad (16)$$

*2.3. Particle Boltzmann equations*

In the same manner as in H06, we assume that the gap is stationary in the co-rotating frame. In this case, it is convenient to normalize the electronic and positronic distribution functions, $N_-$ and $N_+$, per unit magnetic flux. Then, $n_\pm \equiv N_\pm/(\Omega B/2\pi ce)$ obey the following Boltzmann equations:

$$c\cos\chi\frac{\partial n_\pm}{\partial s} + \frac{dp}{dt}\frac{\partial n_\pm}{\partial p} + \frac{d\chi}{dt}\frac{\partial n_\pm}{\partial\chi} = S_\pm, \quad (17)$$

where temporal derivatives of momentum $p$, pitch angle $\chi$, and distance $s$ are given by

$$\frac{dp}{dt} = qE_\parallel\cos\chi - \frac{P_{SC}}{c}, \quad (18)$$

$$\frac{d\chi}{dt} = -\frac{qE_\parallel\sin\chi}{p} + c\frac{\partial(\ln B^{1/2})}{\partial s}\sin\chi, \quad (19)$$

$$\frac{ds}{dt} = c\cos\chi. \quad (20)$$

The collision terms are included in $S_+$ and $S_-$; particles are created by one-photon and two-photon pair productions, and change momenta by inverse-Compton scatterings. The explicit expressions of $S_+$ and $S_-$ are given by equations (36)-(44) in H06. Since the energy transfer in an emission of a single curvature photon is too small to resolve with the current energy grids, we take into account the synchro-curvature effects in the right-hand side as the radiation-reaction force, $P_{SC}$/c; its explicit form is given by Chang & Zhang (1996), or equations (32)-(35) in H06. For relativistic particles, the Lorentz factor is given by $\Gamma = p/(m_e c)$.

*2.4. Photon Boltzmann equation*

Since the gap is assumed to be stationary in the co-rotating frame, the Boltzmann equation for photons becomes (H06)

$$\left(c\frac{k^\varphi}{k} - \Omega\right)\frac{\partial g}{\partial\phi} + c\frac{k^r}{k}\frac{\partial g}{\partial r} + c\frac{k^\theta}{k}\frac{\partial g}{\partial\theta} = S_\gamma, \quad (21)$$

where $\phi = \varphi - \Omega t$ and $k \equiv (-k^i k_i)^{1/2}$. The photon propagation $(r,\theta, k^r,k^\theta)$ in the curved spacetime can be computed by geometrical optics (see section 2.4 in H06) if we give the initial photon propagation direction at the emission point. We could assume that photons are emitted along the local field lines in the co-rotating frame and their propagation direction changes due to aberration. However, since there is relative acceleration between the co-rotating and static frames, applying special relativity becomes bad in the outer magnetosphere and completely breaks down outside the light cylinder. Thus, in this paper, we follows the argument of relativistic plasma flows and adopt (Mestel et al. 1985, Camenzind 1986a,b)

$$u^r = \kappa B^r, \quad (22)$$

$$u^\theta = \kappa B^\theta, \quad (23)$$

$$u^\varphi - \Omega u^t = \kappa \frac{-B_\varphi}{c^2 \rho_w^2}, \quad (24)$$

where the four velocity $u^\mu$ is related with the observed velocity $dx^\mu/dt$ by $u^\mu = u^t(dx^\mu/dt)$. Since the definition of the proper time, $u^\mu u_\mu = c^2$, gives $u^t$, we can compute the photon emission direction observed by a distant static observer by the ratios, $u^r/u$, $u^\theta/u$, and $u^\varphi/u$, where $u = (-u^i u_i)^{1/2}$.

*2.5. Boundary conditions*

To solve the set of partial differential equations (8), (17), and (21) for $\Psi$, $n_+$, $n_-$, and $g$, we must impose appropriate boundary conditions. First, let us consider the elliptic type equation (8). For convenience, we put $\Psi=0$ on the NS surface, that is, the inner boundary. We assume that the lower boundary coincides with the last-open field lines, $\theta_* = \theta_*^{\max}(\varphi_*)$, which is grounded to the star; thus, we also impose $\Psi=0$ on the lower boundary. Since pair production is copious above a certain height from the last-open field line, and since $E_\|$ is highly screened by the discharge of the produced pairs there, we can safely assume that $\Psi=0$ also holds above a certain height, $\theta_* = (1-h_m)\theta_*^{\max}(\varphi_*)$. If we take a larger $h_m$, co-latitudinal grid resolution becomes bad. If we take a smaller $h_m$, on the other hand, there is a risk of excluding active field lines from consideration. For the Crab pulsar, we find that $h_m = 0.26$ is a good compromise. The results change little if we increase (or decrease) $h_m$ 30% from this value. We parameterize magnetic co-latitudes with

$$h \equiv 1 - \theta_* / \theta_*^{\max} \quad (25)$$

and solve the gap in $0 < h < h_m$, where $h=0$ gives the last-open field line. The outer boundary is defined as the place where $E_\|$ changes sign near the light cylinder and found to appear near the place where $\partial(\rho_{GJ}/B)/\partial s$ vanishes due to the curving up of the field lines towards the rotation axis (see discussion around eq. [67] in H06 for physical reasoning).

Next, let us consider the hyperbolic type equations (17) and (21). In the same manner as in the 2-D analysis, we assume in this 3-D examination that electrons, positrons and photons do not penetrate into the gap across the inner and outer boundaries. However, if the created current becomes greater than the Goldreich-Julian value, a positive $E_\|$ is maintained at the NS surface, which leads to an extraction of ions from the surface as a SCLF (see section 2.5 of H06 for details). Even though these ions do not efficiently radiate, they contribute to the charge density in the outer magnetosphere and reduce $E_\|$ to some extent; this contribution is included in the real charge density $\rho$ in the Poisson equation (8). Since we are going to consider copious radiation, pair production, and global electric current flows, we do not speculate the possibility of the formation of the 'electrosphere' as suggested by Krause-Polstorff & Michel (1985) and Wada & Shibata (2007). Throughout this paper, we consider a rotator in the sense that the magnetic axis resides in the same hemisphere as the rotation axis, that is, $\vec{\Omega}\bullet\vec{B} > 0$. In this case, $\rho_{GJ}<0$ holds in the PC region.

## 3. Results: Outer-gap solution

In this section, we apply the method to the Crab pulsar. As representative values, we adopt $\alpha=60^\circ$, $kT$=100 eV, and $\mu=4\times10^{30}$ G cm$^3$. Since we are motivated to elucidate the gap electrodynamics in this paper, we do *not* adjust these parameters to obtain better fits with observations.

*3.1. 3-D gap geometry*

Let us begin by presenting the solved gap geometry in the 3-D magnetosphere. Since $E_\|$ decreases towards the convex side of the curved field lines (i.e., with increasing $h$) relatively gradually, we can practically define $h_m$, above which gap activity is negligible. To grasp how the gap activity decreases at larger $h$, let us first present the potential drop along each field line in figure 3, which represents the potential drop as a function of the gap height $h$ measured from the last-open field line, and of the magnetic azimuth $\varphi_*$. The potential drop takes the maximum value $\Psi_{\max}\sim1.4\times10^{15}$ V in the leading side (in this case, $\varphi_*\sim1$ rad) due to less efficient pair production there. The figure tells us that the potential drop decreases enough around $h_m$=0.16.

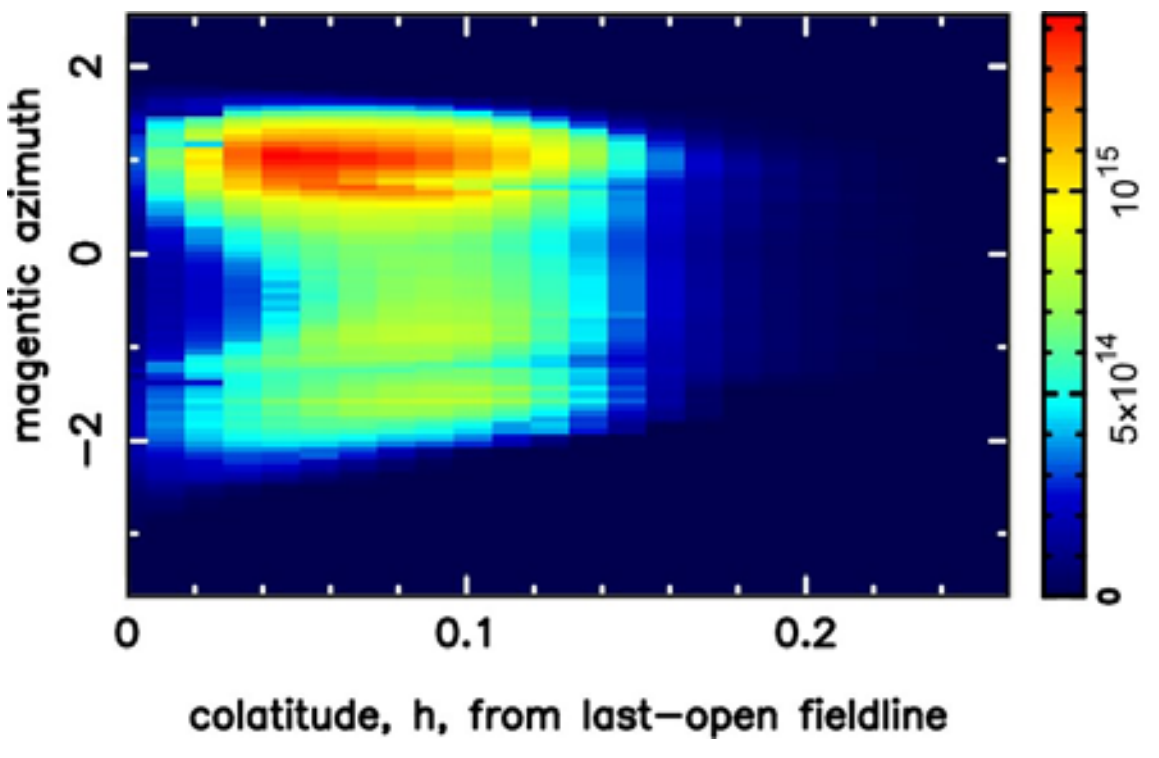

Fig. (3) Distribution of the electrostatic potential drop [V] along the field line that thread the NS surface at magnetic colatitude $h$ and azimuth $\varphi_*$.

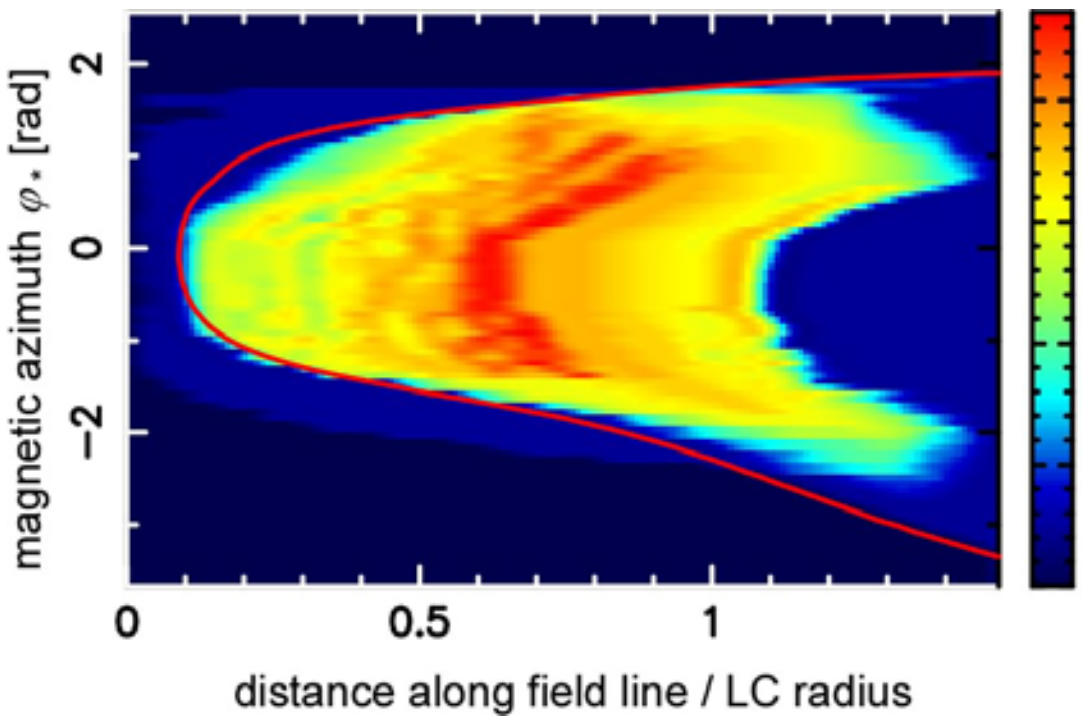

Fig. (4) Gap trans-field thickness on the last-open field line bundle. The red curve denotes the intersection with the null surface where $\rho_{GJ}$ vanishes.

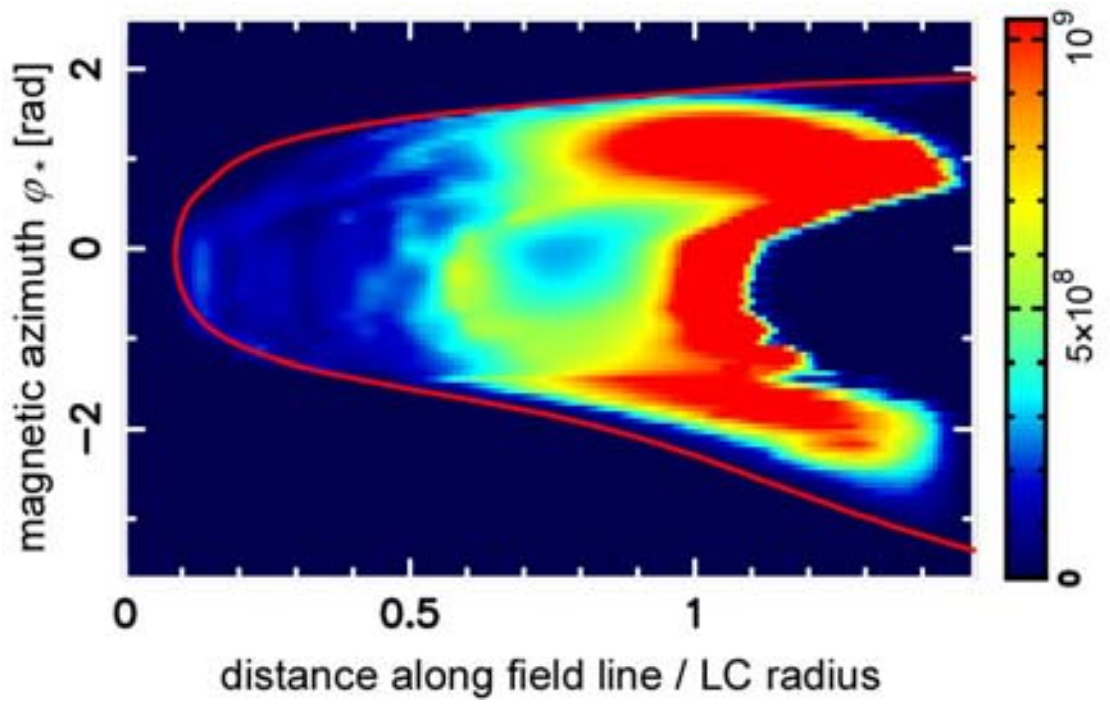

Fig. (5) Distribution of the maximum $E_\parallel$ [V m$^{-1}$] near the gap center (in the meridional direction) at each point on the 2-D last-open field line surface ($s$,$\varphi_*$).

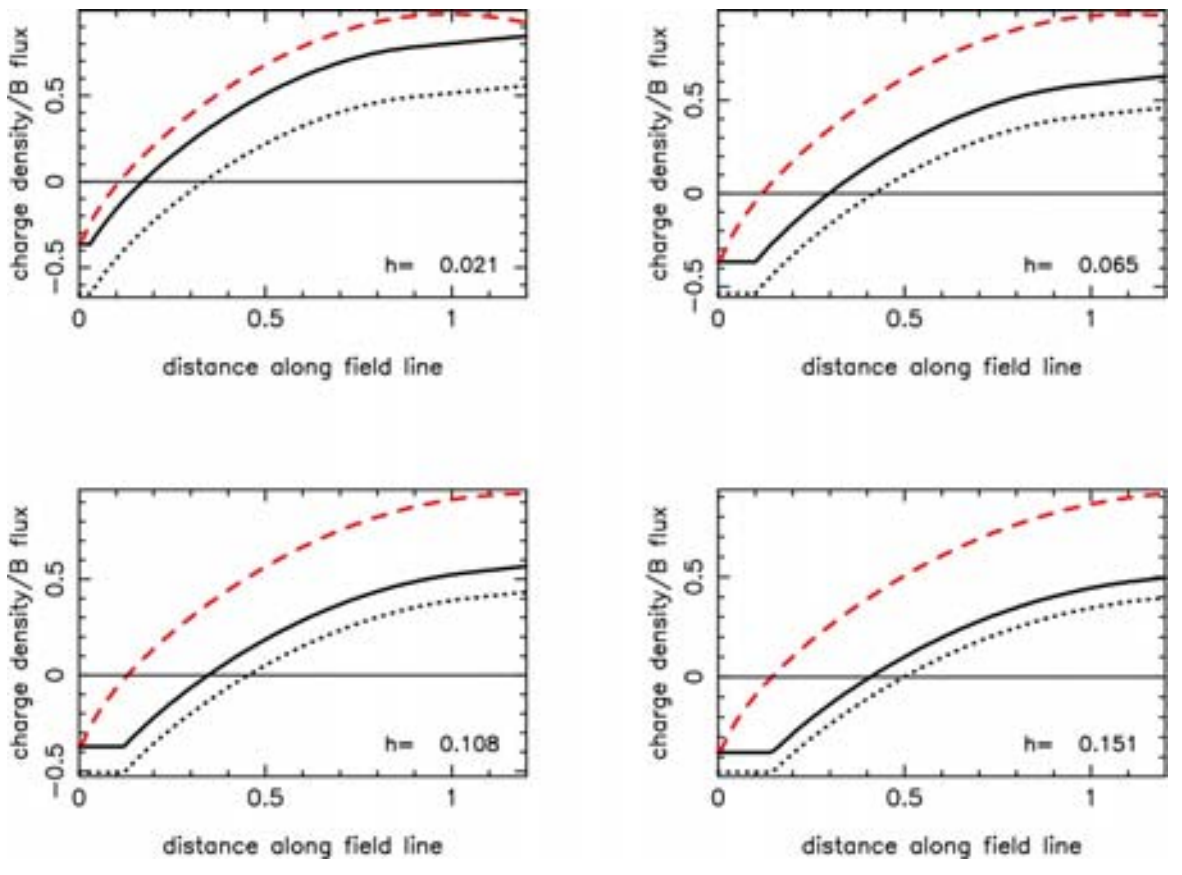

Fig. (6) Total (solid), created (dotted), and Goldreich-Julian (dashed) charge densities in ($\Omega B/2\pi c$) unit, for $\varphi_*$=-30$^o$ at four different co-latitude, $h$. The solid curve includes the ionic contribution, while the dotted curve represents that from the electrons and positrons only.

To quantify $h_m$, we define $h_m$ above which $E_\parallel$<0.01 $\Psi_{max}/\varpi_{LC}$ holds. The gap thickness defined in this way, distributes on the last-open field line surface (i.e., $s$-$\varphi_*$ plane), as shown in figure 4. It shows that the gap is active outside the null surface, as originally suggested by Cheng et al. (1986a). The obtained gap thickness, $h$~0.2 is consistent with phenomenological OG models (e.g., Tang et al. 2007).

### *3.2. Acceleration electric field*

The magnetic-field-aligned electric field, $E_\parallel$, maximizes near the center ($h$~0.08) of the red/yellow 'island' in figure 3. This maximum $E_\parallel$ at each ($s$,$\varphi_*$) is depicted in figure 5. Since photons propagate into smaller $\varphi_*$ direction (i.e., propagate clockwise in the co-rotating frame), unless the field lines become mostly toroidal (as expected outside the light cylinder), pairs are preferentially produced in the trailing side (i.e., at smaller $\varphi_*$). As a result, $E_\parallel$ is screened and hence potential drop decreases (fig. 3) in the trailing side, making the second peak weaker than the first peak (section 3.3). Figure 5 also shows that $E_\parallel$ is substantially screened in the inner part of the gap. This is because the produced pairs discharge so that the original $E_\parallel$ may be reduced.

In figure 6, we present the charge density per magnetic flux tube, $\rho/(\Omega B/2\pi c)$, as the solid curve, together with $\rho_{GJ}/(\Omega B/2\pi c)$ as dashed. It follows that $\rho/B$ has a similar gradient with $\rho_{GJ}/B$; this result is consistent with the analytical condition (section 3.3 in H06) so that $E_\parallel$ is screened most efficiently. It is worth noting that ions could be extracted as a SCLF from the NS surface if the produced current exceeds the GJ value at the NS surface (section 3.3 in H06). In this case, the real charge density increases from the dotted curve to the solid one, where the former only includes the contribution of only the electrons and positrons produced in the gap.

### *3.3. Photon mapping result and pulse profile*

We plot the photon intensity on the pulse phase $\Phi$ vs. observer's viewing angle $\zeta_{obs}$ plane in figure 7. The photons emitted from the north magnetic pole would appear at $\Phi$=0 and $\zeta_{obs}=\alpha$ (=60$^o$). The left panel shows that the OG connected to the north pole emits outward-propagating photons in $\zeta_{obs}$ >90$^o$, and -140$^o$<$\Phi$< 40$^o$, while inward ones in $\zeta_{obs}$ <90$^o$ and 130$^o$<$\Phi$< 300$^o$ (=-60$^o$). On the contrary, from the south-pole OG, inward-propagating photons appear in $\zeta_{obs}$ >90$^o$ and -50$^o$<$\Phi$< 120$^o$. Thus, the intensity around $\zeta_{obs}$ ~120$^o$ and $\Phi$~ 0$^o$ consists of both the outward emission from the north-pole OG and the inward emission from the south-pole OG (right panel). This result is consistent with phenomenological 3-D OG models (Takata et al. 2007; Tang et el. 2008) and self-consistent 2-D OG solution (Takata et al. 2008).

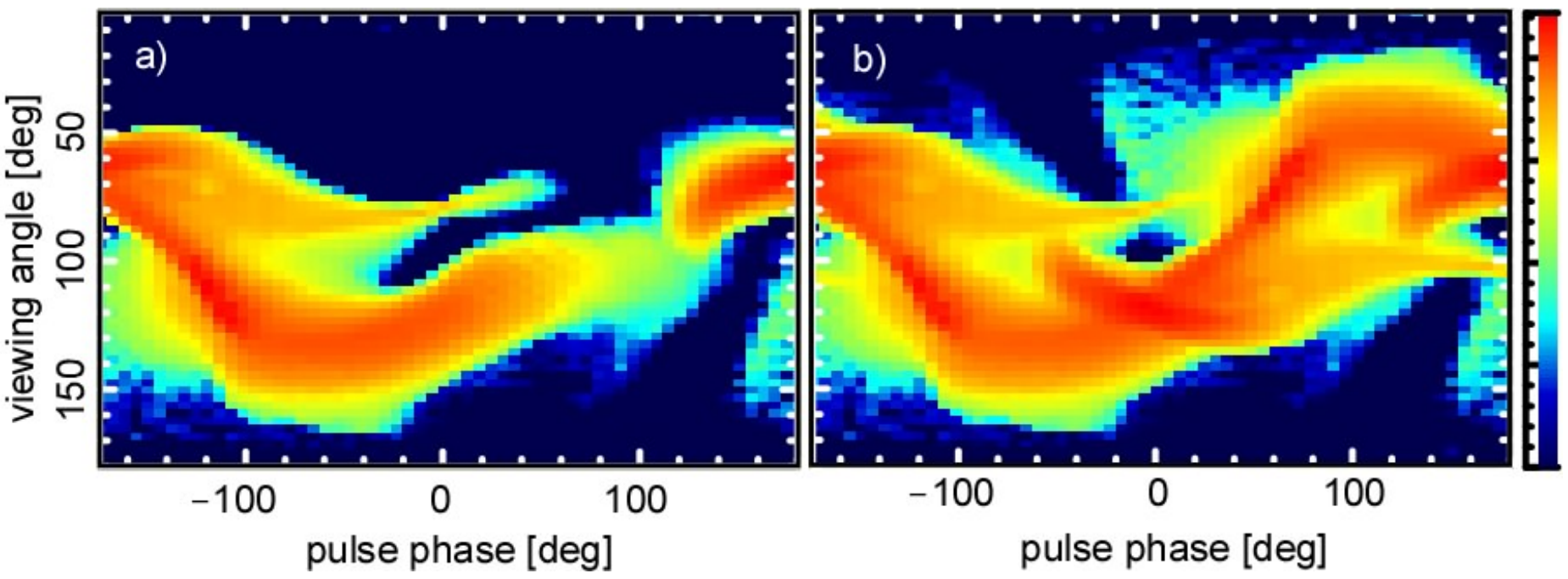

Fig. (7) Phase plot of the photons having energies above 90 MeV. *Left:* Emission from the north-pole OG alone; *right:* emission from the both poles. Reddened regions indicate strong intensity.

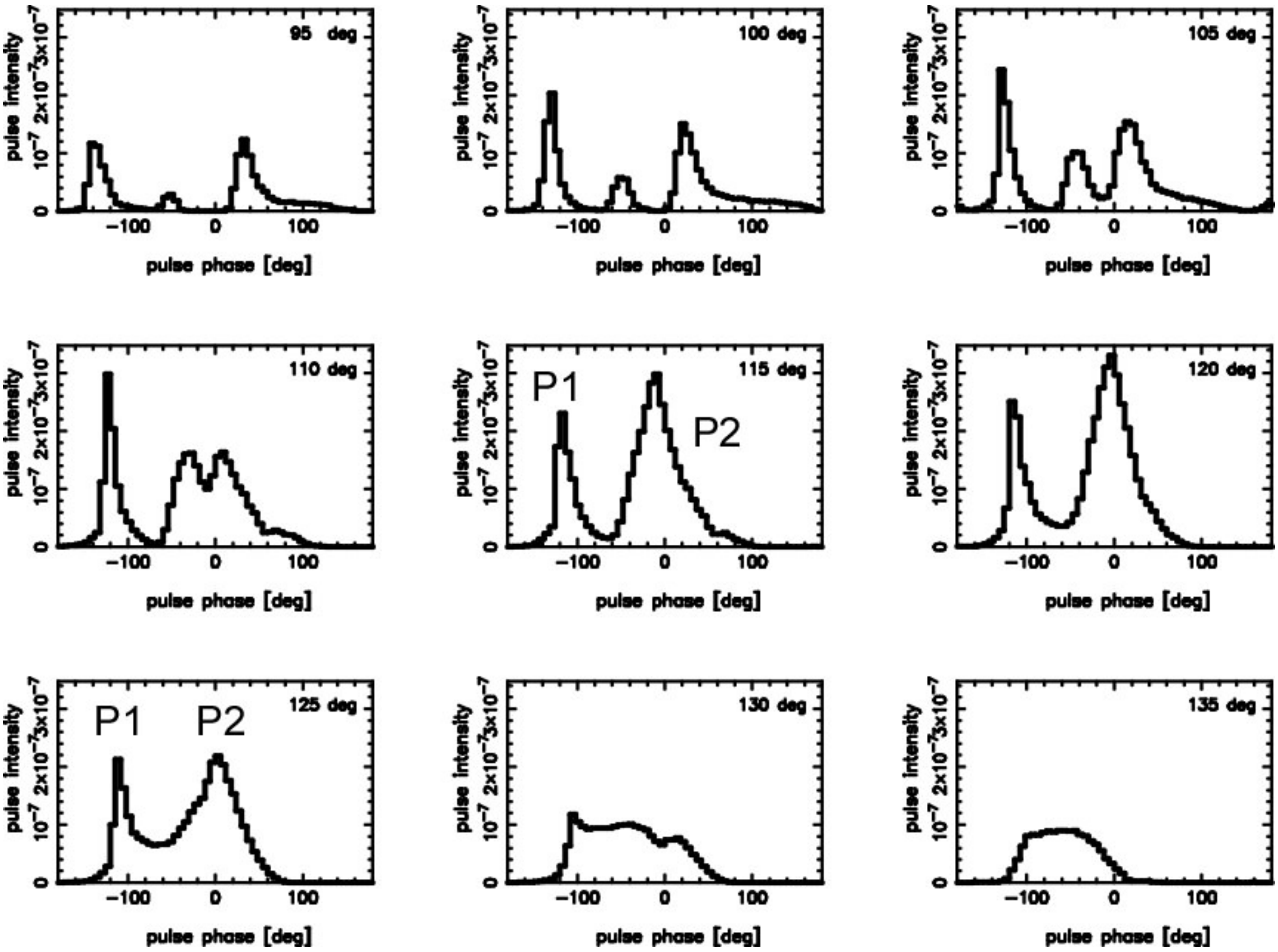

Fig. (8) Pulse profile for nine different viewing angles, $\zeta_{obs}$. Normalization is common in all the panels.

If we superpose the south-pole SG emissions and specify $\zeta_{obs}$, we obtain the pulse profile as presented in figure 8, which demonstrates how the profile depends on $\zeta_{obs}$. Comparing with observations, preferable $\zeta_{obs}$ is found to be in the range $105^o<\zeta_{obs}<125^o$ for $\alpha=60^o$. We should notice here that the first peak (P1) composed of the outward emission from the outer-most part of the OG, of which distance from the rotation axis is $\varpi>0.7\,\varpi_{LC}$. Thus, the shape, strength, and profile of P1 may have to be revised if we incorporate the magnetic field deformation due to magnetospheric currents near the light cylinder, instead of giving the magnetic field configuration from the current-free, rotating magnetic dipole formula as in the present work.

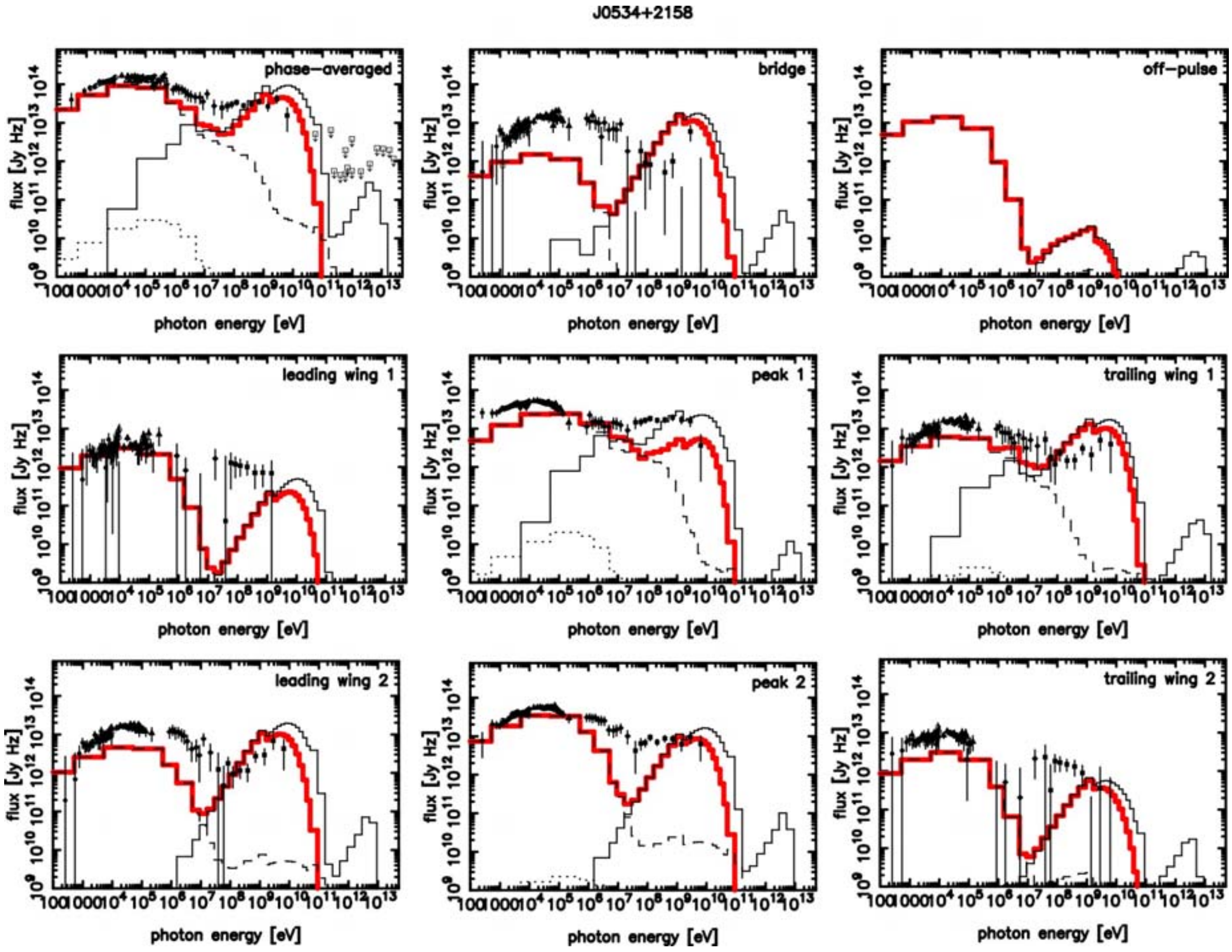

Fig. (9) Phase-averaged (top left) and phase-resolved (others) spectrum of the pulsed emission from the Crab pulsar. The thin solid, dashed, and dotted curves denote the un-absorbed photon fluxes of primary, secondary, and tertiary components, respectively, while the thick red solid one presents the spectrum to be observed, including absorption along the line of sight. Interstellar absorption is not considered. The filled circles (LECS), open circles (MECS), filled triangles (PDS) denote the Boppo SAX observations, while the open triangle the Gamma-ray Imaging Spectrometer (GRIS), and inverse filled triangles (OSSE), diamonds (COMPTEL), and filled squares (EGRET) denotes the CGRO observations. Data points are from Kuiper et al. (2001) [http://www.sron.nl/divisions/hea/kuiper/data.html]. In the top left panel, upper limits by ground-based observations are also plotted above 50 GeV.

### *3.4. Radiation spectrum*

Let us finally examine the radiation spectrum. From figure 8, we adopt $\zeta_{\rm obs}=125^{\rm o}$ as an appropriate value to obtain the phase-resolved spectrum as presented in figure 9. It follows that the predicted spectrum is roughly consistent with observations. At P1 (in pulse phase $40.4^{\rm o}-76.4^{\rm o}$, central panel), absorption suppress the flux above 50 MeV, as we can see from the un-absorbed primary flux (black thin solid curve) and the final flux to be observed (red thick solid curve). This is because most of the photons are emitted near the light cylinder ($\varpi \sim 0.8 \varpi_{\rm LC}$) and hence have large 3-D collision angles (>0.2 rad) with the strong magnetosphereic X-ray photons owing to the curving up field lines towards the rotation axis. In other pulse phases, absorption becomes important typically above a few GeV. At P2 ($177.2^{\rm o}-216.8^{\rm o}$, bottom center), the synchro-curvature photons emitted from the primaries in a residual, small-amplitude $E_\parallel$ region (see the region $\varphi_* \sim -2$ rad in fig. 5) is less sufficient, resulting in the insufficient flux around 10 MeV. The off-pulse emission ($249.2^{\rm o}-18.8^{\rm o}$) appears to be small from the pulse profile (fig. 8) but not negligible in the phase-resolved spectrum (fig. 9). This is because outward photons appear after P2 being emitted close to or beyond the light cylinder by curvature radiation and ICS of the magnetospheric IR photons (Aharonian & Bogovalov 2003), and because inward photons appear around phase $20^{\rm o}$ (for $\zeta_{\rm obs}$ $\sim 120^{\rm o}$) being emitted from the south-pole OG. The ICS photons are also emitted in other phases; however, they are totally absorbed by the same IR photon field to appear as the secondary synchrotron component between 100 MeV and 100 GeV, which is negligibly small compared with the primary curvature flux. Since $E_{\parallel}$ is substantially screened by the produced pairs in the inner part of the OG (e.g., at $s$<0.5 $\varpi_{\rm LC}$), the outward emission from the north-pole OG does not contribute for the bridge (112.4°-152.0°, top center) very much. Instead, the inward emission from the south-pole OG contributes to the bridge flux to some extent. The data points are from Harnden & Seward (1984), Pravdo & Serlemitsos (1981), Knight (1982), Weisskopf et al. (2004), and Mineo et al. (2006) for X-rays; from Nolan et al. (1993), Ulmer et al. (1995), Much et al. (1995), Fierro et al. (1998), Kuiper et al. (2001) for 10~MeV-20GeV; from Borione et al. (1997), Tanimori et al. (1998), Hillas et al. (1998), Lessard et al. (2000), and de Naurois et al. (2002) for the upper limits above 50 GeV.

## 4. Slot-gap model

We apply the same numerical scheme described in sections 2 and 3 to the SG model. In section 4.1, we briefly review how the SG is formed in the PC region. Since nobody has independently re-examined the SG model so far, we first check the SG model (MH04; MH04; HSDF08) by our code in sections 4.2-4.3.

### *4.1. Formation of a slot gap in polar cap region*

For a parallel rotator, $\vec{\Omega}\bullet\vec{B}>0$, in a PC space-charge-limited flow (SCLF) model, electrons are extracted at a slightly smaller rate (in an absolute value sense) than the GJ rate. We may write this electric current density as $j=\rho c=(1-\delta)c\rho_{\rm GJ}^{*}$, where $\delta$ is a small positive constant and * denotes that the quantity is evaluated at the NS surface. Provided that the magnetosphere is stationary in the co-rotating frame and that pair production does not take place, $j/B \propto \rho/(\Omega B/(2\pi c)$ conserves along the field line, as indicated by the green dotted line in figure 10. On the other hand, along the magnetic field lines curving away from the rotation axis, the Newtonian GJ charge density per magnetic flux tube, $\rho_{\rm GJ}/(\Omega B/(2\pi c)$ $=-B_{\zeta}/B$, where $B_{\zeta}$ denotes the magnetic component projected along the rotation axis, increases outwards, as indicated by the black thin dashed line. Thus, at the altitude much less than the stellar radius, $\rho_{\rm eff} \equiv \rho_{\rm GJ}-\rho$ changes sign from negative to positive outwards, leading to a sign reversal of $E_{\parallel}$ from negative to positive, thereby terminating the gap. (Exactly speaking, the place where $E_{\parallel}$ changes sign depends not only on the sign of $\rho_{\rm eff}$ but also on the trans-field structure; nevertheless, this argument of gap termination is qualitatively correct in any case.) Thus, it had been considered that a PC accelerator is non-active along the polar field lines curved away from the rotation axis, localizing in the vicinity of the NS surface (Arons 1983).

Subsequently, MT92 pointed out that an energetic PC accelerator can exist also along away-curvature field lines due to general relativistic (GR) effects, because $\rho_{\rm GJ}/B$ deviates from the Newtonian value due to the $\Omega$-$\omega$ factor (instead of $\Omega$) in the right-hand side of equation (9), as indicated by the black thick solid curve in fig. 10. Since the SCLF has $\rho/(\Omega B/2\pi c)$ as indicated by the red thick solid curve for this revised PC model, a negative-definite $\rho_{\rm eff}$, which ensures a negative-definite $E_{\parallel}$, extends up to a greater altitude than the Newtonian case. Thus, a PC accelerator exists within the altitude comparable to the NS radius for a rapidly rotating NS as in the Crab pulsar system. The resulting potential drop is found be enough to reproduce high energy emissions (for details, see MH03 and references therein).

In spite of these attractive features on energetics, the GR-PC accelerator is still localized within a few stellar radii even for millisecond pulsars. Thus, the solid angle is too small to reproduce the wide pulsed profiles that are observed. They were, therefore, motivated by the need to contrive a higher altitude emission model.

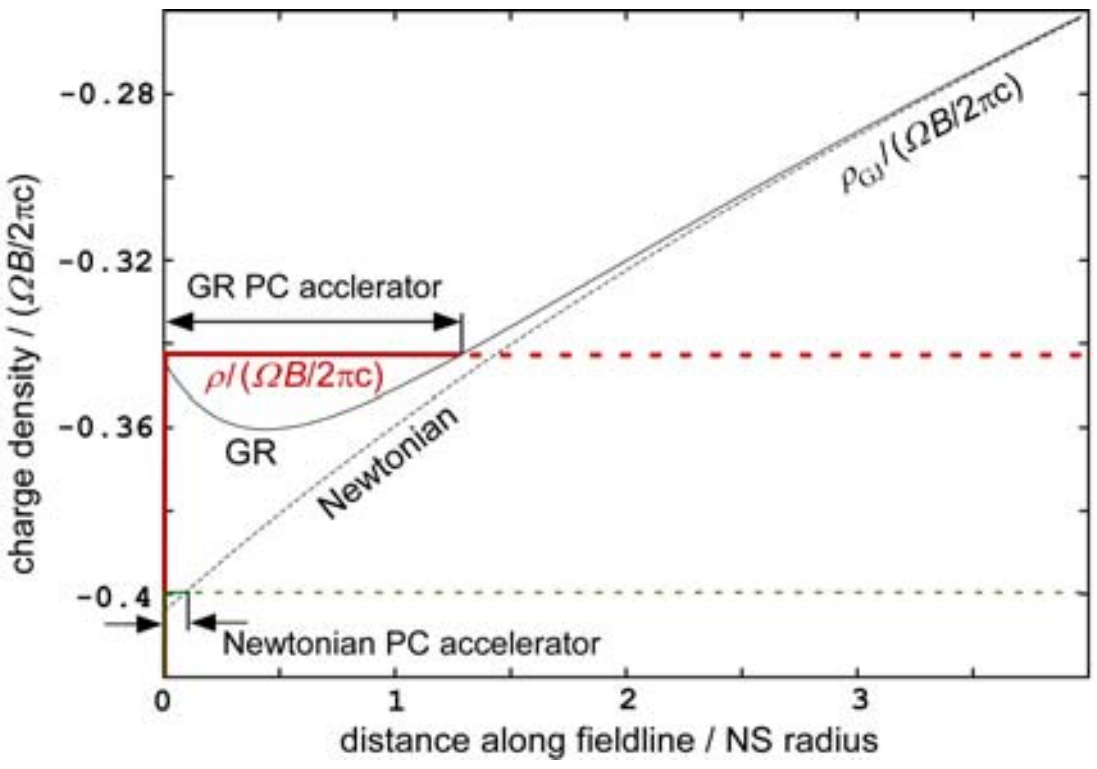

Fig. (10). Charge density per magnetic flux tube along the field line curved away from the rotation axis. The black solid curve denotes the Goldreich-Julian charge density (i.e., eq.[9]) for a dipole magnetic field with GR corrections, while the black dashed one the Newtonian value. The rotational frequency, radius, and magnetic inclination of the NS are 33 ms, 10km, and 60°, respectively. The magnetic field line is selected to be the last-open field line that cross the NS surface at $\varphi_{*}$=0. The green thin solid curve (bottom left corner) denotes the real charge density for Newtonian PC model, while the red thick solid one for the PC model with GR corrections.

Developing the idea that the pair formation front (PFF), above which $E_{||}$ is screened, occurs at increasingly higher altitude as the magnetic colatitude approaches the last-open field line where $E_{||}$ vanishes (Arons & Scharlemann 1979), MH03 first investigated a self-consistent electrodynamics of the SG, a narrow space between PFF (upper boundary) and the last-open field line (lower boundary), and found that $E_{||}$ is greatly reduced by the proximity of the two conducting boundaries.

### *4.2 Lower-altitude solution*

We re-examine the SG model of MH03 by applying our numerical code. MH03 demonstrated that the PFF becomes almost parallel to the NS surface in the central part of the PC (i.e., near the magnetic axis) and that the PFF rapidly curves up to become almost parallel to the last-open field line above several per cent of $r_*$ for a rapidly rotating, highly magnetized NS. We confirmed their results by our code as indicated by figure 11, which presents the height, $h_m$, of the PFF measured in meridional direction from the last-open field line as a function of $s$ and $\varphi_*$. It is also confirmed that the potential drop (fig. 12) is consistent with MH03 (see their eq. [22]), being ~$1.5\times10^{12}$V along the field lines threading the higher-altitude SG (e.g., at $h$=0.02, see next subsection for details on the high-altitude SG). Close to the star, the SG structure is found to be almost independent of $\varphi_*$. The electrons are accelerated up to the Lorentz factor of ~$3\times10^6$ at s~$0.08r_*$.

### *4.3 Higher-altitude solution*

The SG solution described just above exists only within the region in which $\rho_{\rm eff}=\rho_{\rm GJ}-\rho<0$ holds to guarantee a negative-definite $E_{||}$ throughout the gap. Therefore, if $\rho/B$ is conserved along the field line (as indicated by the horizontal red solid line in fig. 10), the pair-free SG extends only up to the altitude 1.3 $r_*$ in the case of the Crab pulsar. However, if the SG extends only up to such an altitude, the problem of the small emission solid angle occurs in the same way as in PC models. Thus, MH04 examined the possibility of an extension of the SG into the higher altitudes. However, unlike the OG solution, the higher-altitude SG model cannot be constructed without additional assumptions. We thus describe them in sections 4.3.1 and 4.3.2, then turn to a re-examination of the high-altitude SG model of MH04 and HSDF08 in sections 4.3.3 and 4.3.4.

#### *4.3.1 Additional assumption: real charge density*

To extend the SG into the higher altitudes, MH04 assumed that $\rho/B$ increased with increasing $s$, as indicated by the red solid curve in figure 13. Their rationale is that the SG physical thickness is less than $0.4r_*$ even at the light cylinder (for $\Delta\xi_{\rm SG}=h_m$~0.04 in their notation) and that the cross-field motion of charges would supply enough positive charges that realizes the outward increase of $\rho/B$ in this thin gap (e.g., Luo & Melrose 2007 for trapping of relativistic plasmas in the closed zone). However, this is, unfortunately, not the case. For example, the Newtonian dipole field geometry,

$$\frac{\sin^2(\theta-\alpha)}{r}=\frac{\sin^2\theta_*}{r_*}\ , \qquad (26)$$

gives the gap physical thickness $rd\theta=2.3(h_m/0.04)r_*$ at $r=0.5\varpi_{\rm LC}$ and $6.4(h_m/0.04)r_*$ at $r=\varpi_{\rm LC}$ for $\varpi_{\rm LC}=160r_*$ (i.e., for the Crab pulsar). For cross-field motion to effectively work, the thickness should less than $0.05r_*$ (see section 4.5 in H07); therefore, it is difficult to justify the $\rho/B$ variation (red solid curve) with cross-field motion, provided that $h_m>0.001$.

On these grounds, we abandon to physically justify the $\rho/B$ variation and simply follow the assumption made by equations (28)-(30) in MH04, which ensures the negativity of $\rho_{\rm eff}=\rho_{\rm GJ}-\rho$, and hence $E_{||}$, from the PC region to the higher altitudes. In our notation, it is equivalent with adopting (eq.[38] in MH04)

$$\frac{\rho_{\rm eff}}{B}=-\frac{\Omega}{2\pi c}\left\{\left[\kappa\left(\beta-\frac{r_*^3}{\varpi_{\rm LC}^3}\right)+1-\beta\right]\cos\alpha\right.$$
$$\left.+\frac{3}{2}H(r_*)\theta_*^{\rm max}\left[\frac{H(\varpi_{\rm LC})}{H(r_*)}\sqrt{\frac{\varpi_{\rm LC}f(r_*)}{r_*f(\varpi_{\rm LC})}}-\beta\right]\sin\alpha\cos\varphi_*\right\}, \qquad (27)$$

where $\kappa$~0.15 describes the space-time dragging effect, $\beta, H, f$ are close to unity, $\beta=(1-3r/4\varpi_{\rm LC})^{1/2}$, $H(r)$~$1-0.1x-0.026x^2-0.075x^3+0.019x^4$, $f(r)$~$1+0.3x+0.096x^2$, with $x=r_*/r$.

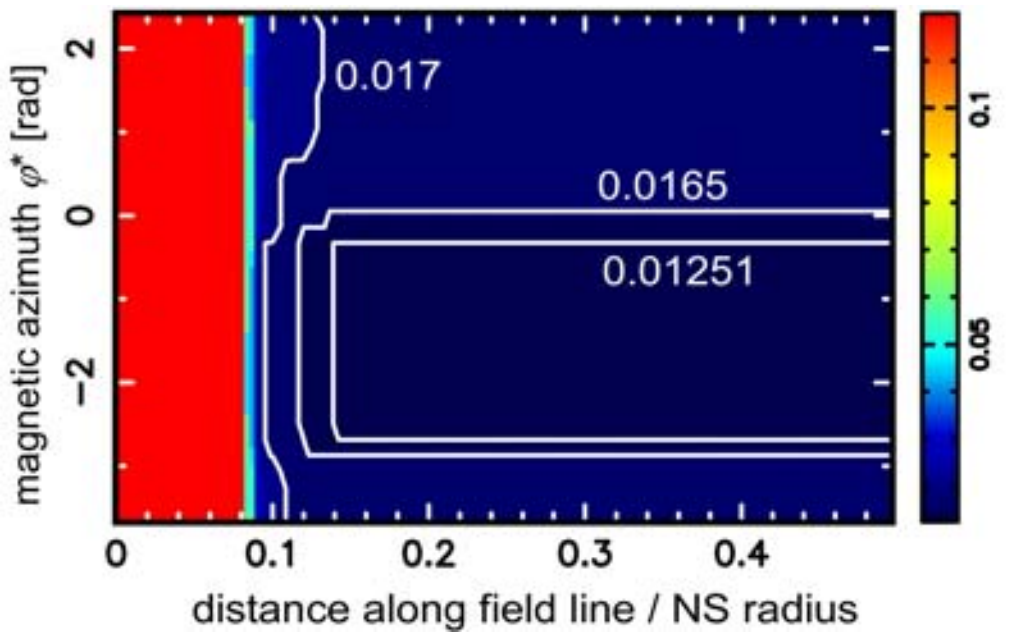

Fig. (11) PFF height, $h_m$ on the last-open field line surface in the lower-altitude SG model. The minimum value is 0.01250. In the reddened region, PFF becomes parallel to the NS surface; thus, the gap exists over the entire PC surface.

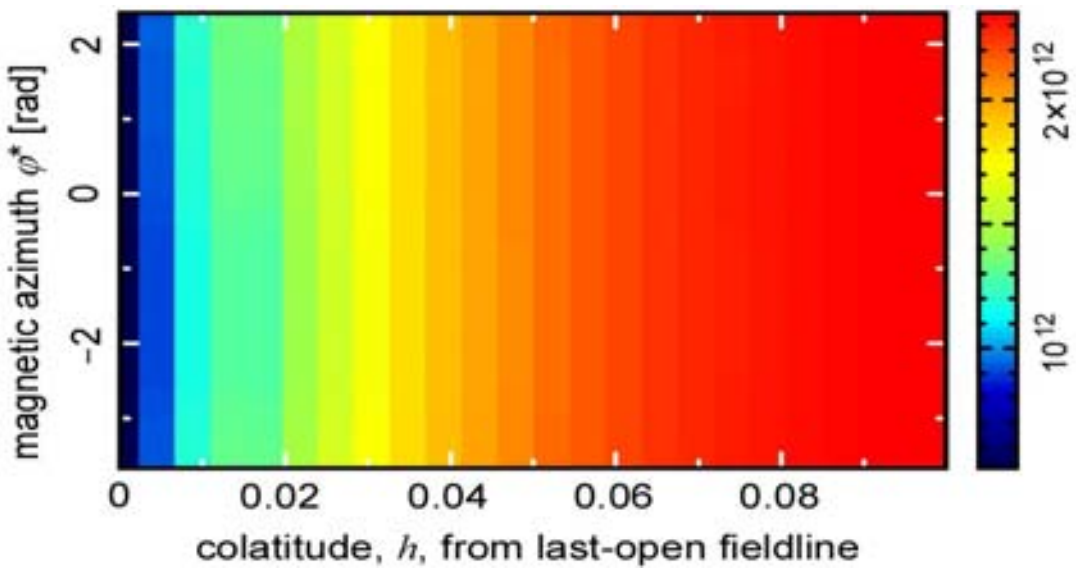

Fig. (12) Electric potential drop [V] along the field line threading the NS at colatitude $h$ and azimuth $\varphi_*$, in the lower-altitude SG model

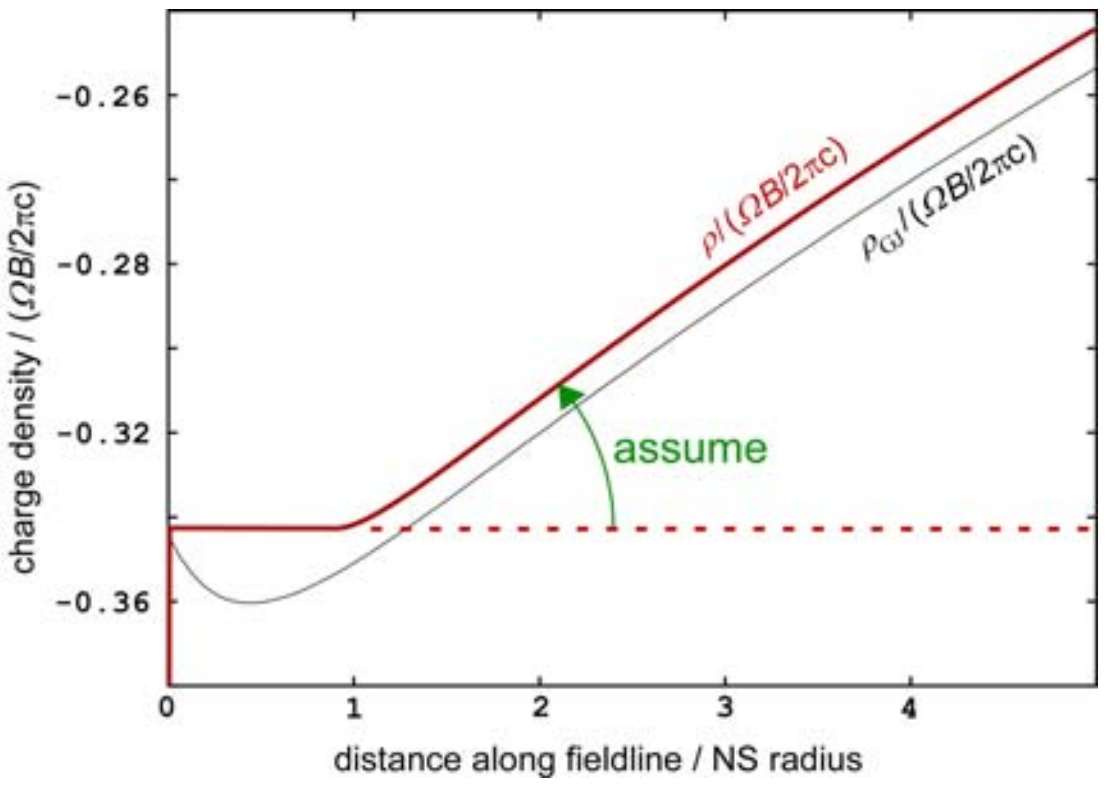

Fig. (13) Schematic picture of the real charge density per magnetic flux tube (red curve). The pair-free assumption of the SG model would conserve the value of $\rho/(\Omega B/2\pi c)$ as the red dashed line. However, to extend a negative $E_{\|}$ to the higher altitudes, $\rho/(\Omega B/2\pi c)$ is artificially assumed to vary as the red solid line in the high-altitude SG model. For explicit functional form of $\rho_{\rm eff}/B=(\rho_{\rm GJ}-\rho)/B$, see eq. (27).

*4.3.2 Additional assumption: trans-field thickness*

Next, let us discuss how the pair-free assumption constrains the co-latitudinal thickness, $h_{\rm m}$. It is clear that there is always a small inward flux of produced positrons in the SG, because they cannot escape from the SG by cross field motion provided that $h_{\rm m}>0.001$. Such inward-migrating positrons emit curvature photons, which cascade into copious pairs in the inner magnetosphere (exponentially with the distance along the field lines). If the produced pairs become a non-negligible fraction of the primary electrons extracted from the NS surface, the pair-free assumption breaks down. Thus, in this paper, we define $h_{\rm m}$ of the higher-altitude SG, by imposing that the electric current carried by the produced pairs does not exceed 10% of the primary electron current, $c\rho_{\rm GJ}$, along each field line. Note that this constraint on $h_{\rm m}$ could not be obtained if we only considered the outward-migrating electrons as in previous SG models.

Under this assumption, the $h_{\rm m}$ distribution on the last-open field line bundle (i.e., $s$-$\varphi_*$ plane) becomes as present in figure 14. If $h_{\rm m}$ increases from the value at some point, the increased pair production could be compensated by the decrease of $h_{\rm m}$ at another point along the same field line. (Therefore, the 'solved' $h_{\rm m}$ in fig. 14 is not unique.) However, since $E_{\|}$ is proportional to $h_{\rm m}^2$, the cascaded pairs rapidly increase with increasing $h_{\rm m}$. Thus, the $h_{\rm m}(s, \varphi_*)$ presented in figure 14 gives a good estimate of the upper bound for the SG to be pair-free. In another word, the electric potential drop along each field line cannot become very large compared with the value derived from the $h_{\rm m}$ presented in figure 14.

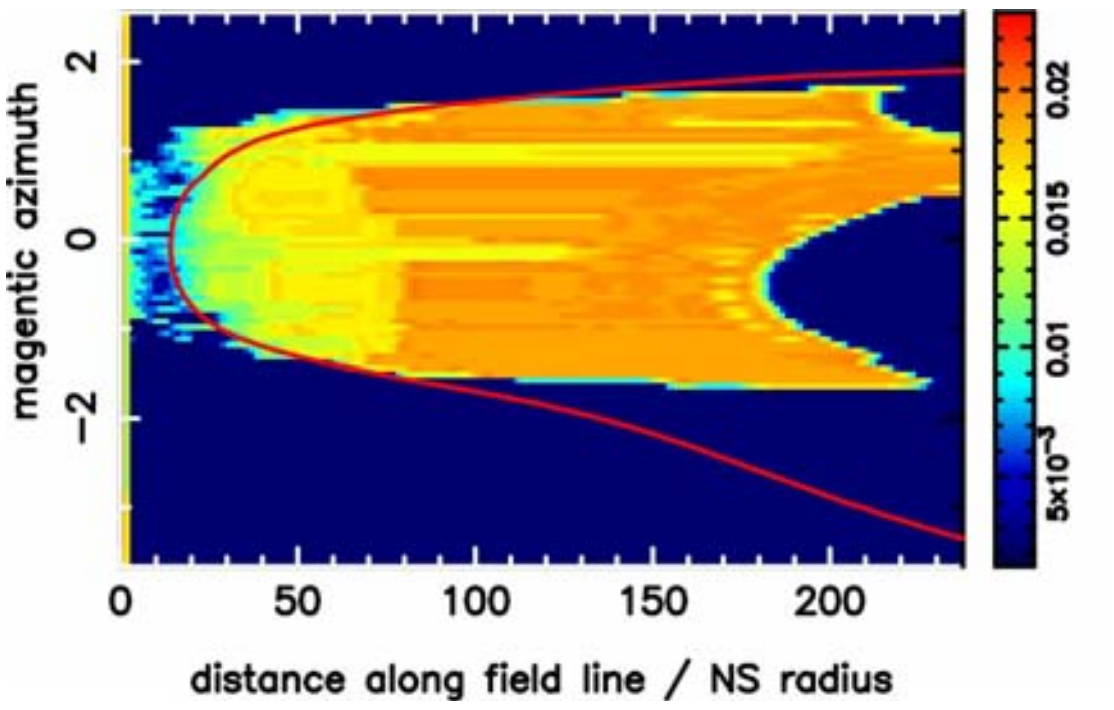

Fig. (14) SG trans-field thickness, $h_{\rm m}$, on the last-open field line surface. The red solid curve represents the position of the null surface.

Note that the gap thickness, $h_{\rm m}$, is constrained by no means from the first principles. It is only artificially imposed so that the SG may be pair-free. It is also noteworthy that $h_{\rm m}$ in the higher altitudes (fig. 14) is less than the minimum PFF height $h_{\rm m}$ (~0.017) solved in the lower altitudes (fig. 11). In the lower altitudes, cascaded pairs above the PFF have outward momenta and hence migrate outwards to screen $E_{\|}$ at higher altitudes than the PFF. Thus, it seems reasonable to suppose that $h_{\rm m}$ in the higher altitudes cannot exceed the value constrained in the lower altitudes. Interestingly, the derived $h_{\rm m}$ (fig. 14) is coincidentally less than or comparable to 0.017, even though we do not artificially set the upper limit of $h_{\rm m}$ from the lower-altitude value. Along the field lines threading the NS surface in $|\varphi_*| > 1.5$ rad, $\rho_{\rm eff}$ becomes positive by equation (27); thus, the SG does not exist in the higher altitudes for this azimuthal range, as indicated by the vanishing gap thickness (i.e., black color) in figure 14.

*4.3.3 Potential drop*

As demonstrated by MH04, the lower-altitude solution (relatively) smoothly match the higher-altitude solution at $s$~$1.2r_*$ for the Crab pulsar. Thus, we assume that the higher altitude solution starts from $1.2r_*$ and the electrons are injected with the initial outward momenta determined by the potential drop obtained in the lower-altitude SG solution (fig. 11). Adopting the assumption described in sections 4.3.1 and 4.3.2, and applying the same numerical code to the higher-altitude SG model, we solve $E_{\|}$, particle distribution functions, and the photon specific intensity. It follows from figure 15 that a strong $E_{\|}$ appears only in the outer-most portions of the SG; therefore, most of the potential drop and hence emission occurs near the light cylinder.

The potential drop in the higher-altitude SG is presented in figure 16. It shows that the potential drop attains only $2\times10^{12}$ V, which is less than 1/5 of HSDF08 prediction, because $h_m$ is constrained to be less than half of their estimate (~0.04) due to the pair-free condition.

### *4.3.4 Emission from the SG*

Using the higher-altitude SG solution, we map the photon intensity as a function of the pulse phase $\Phi$ and the observer's viewing angle $\zeta_{obs}$ in figure 17. The red ring around the north pole (at $\zeta_{obs}=60^o$ and $\Phi=0^o$) indicates that photons are efficiently emitted from relatively lower altitudes (say, $s<0.1\varpi_{LC}$) when electrons lose their initial energy gained in the lower-altitude SG ($s<1.2r_*$). Because $E_{\|}$ is weak in the middle altitudes, caustic emission is negligibly small. (For example, electrons' Lorentz factors attain only $6\times10^6$ at $s$~$0.6\varpi_{LC}$.) However, in the outer-most part of the SG (e.g., $s>0.7\varpi_{LC}$), relatively strong $E_{\|}$ (~$3\times10^6$V m$^{-1}$) leads to the emission represented by the cyan arc around $\zeta_{obs}$~$150^o$ and $-110^o<\Phi<20^o$).

If we superpose the south-pole SG emissions and specify $\zeta_{obs}$, we obtain the pulse profile as presented in figure 18. It follows from the ordinate values in each panel that the photon flux above 90 MeV is a few orders magnitude less than the OG prediction (fig. 8). The strong intensity appearing around $\Phi$~$0^o$ represents the photons emitted relatively close ($r<0.1\varpi_{LC}$) to the south pole.

To calculate the spectrum, we adopt $\zeta_{obs}=60^o$ (or equivalently $120^o$), because the photon mapping result (fig. 17) shows that the photon intensity maximizes at this viewing angle. Figure 19 demonstrates that the predicted flux of the SG emission is negligible from IR to VHE. This is solely because the SG thickness, $h_m$, is severely constrained by the pair-free condition, which has not been considered in any previous SG models.

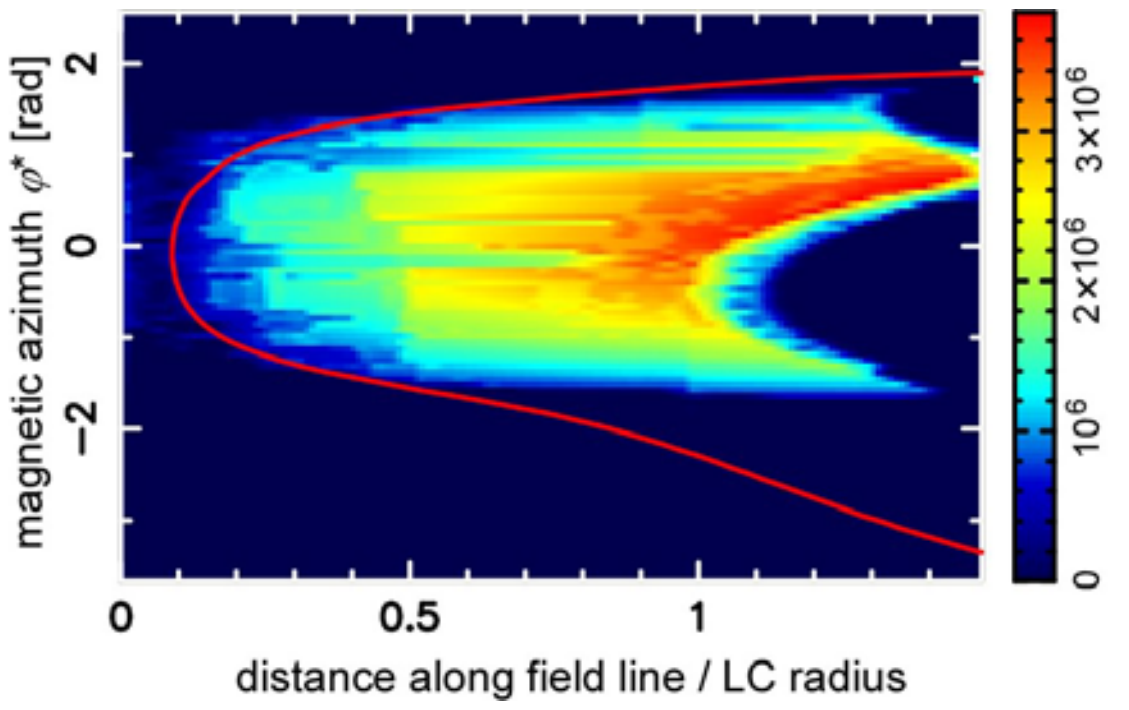

Fig. (15) Acceleration field [V/m] distribution on the last-open field line surface in the higher-altitude SG model. The solid curve represents the null surface position.

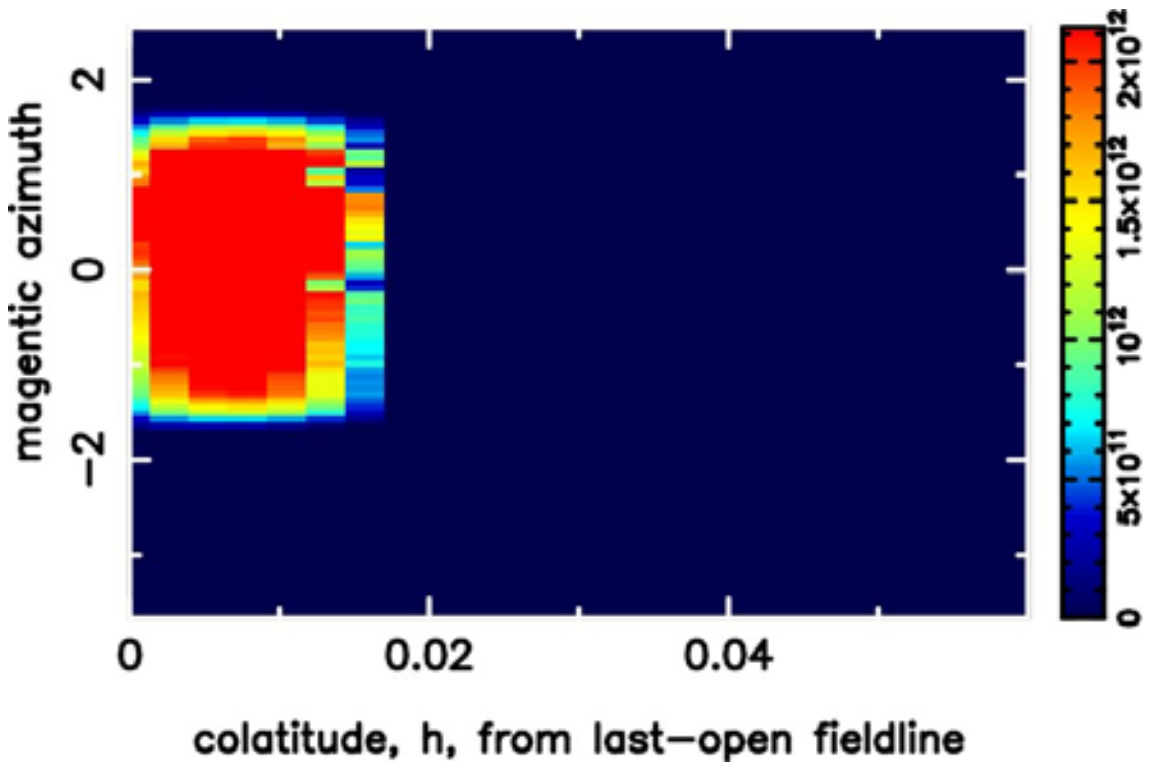

Fig. (16) Electric potential drop [V] along the field line threading the NS at colatitude $h$ and azimuth $\varphi_*$ in the higher-altitude SG model.

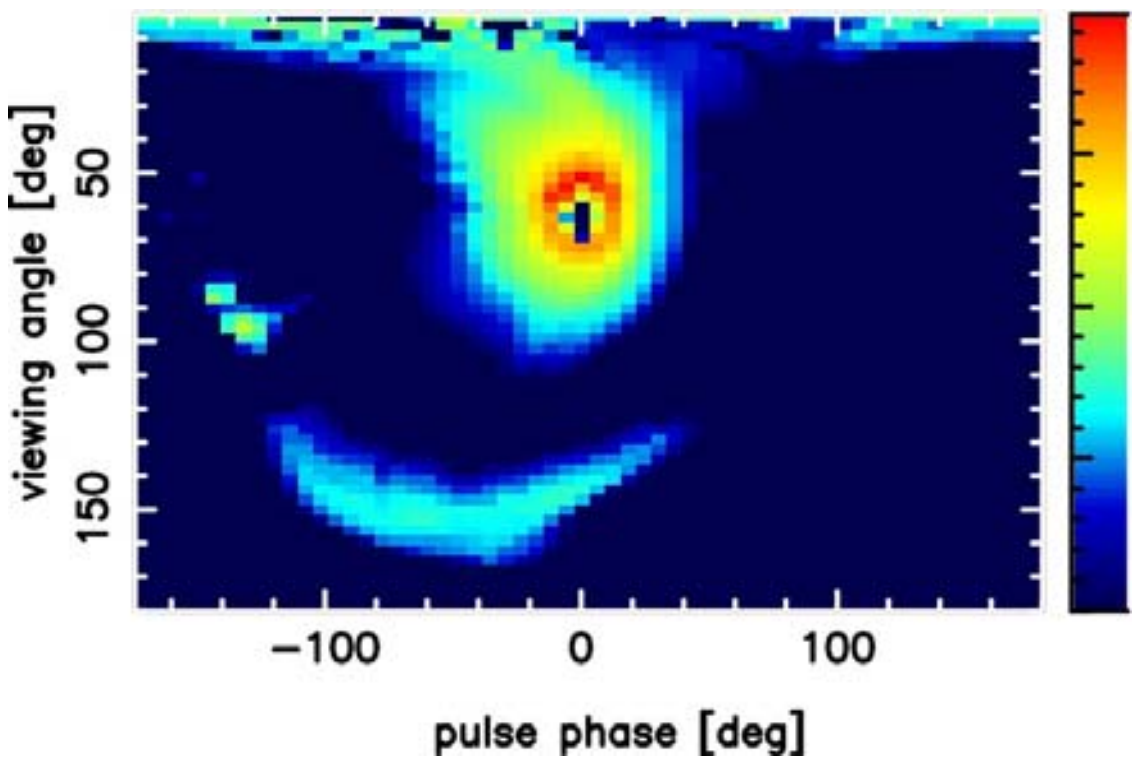

Fig. (17) Photon mapping results on the phase diagram in the higher-altitude SG model. The reddened regions indicate strong photon intensity. Only the photons emitted from the north-pole SG and having energies greater than 90 MeV are plotted.

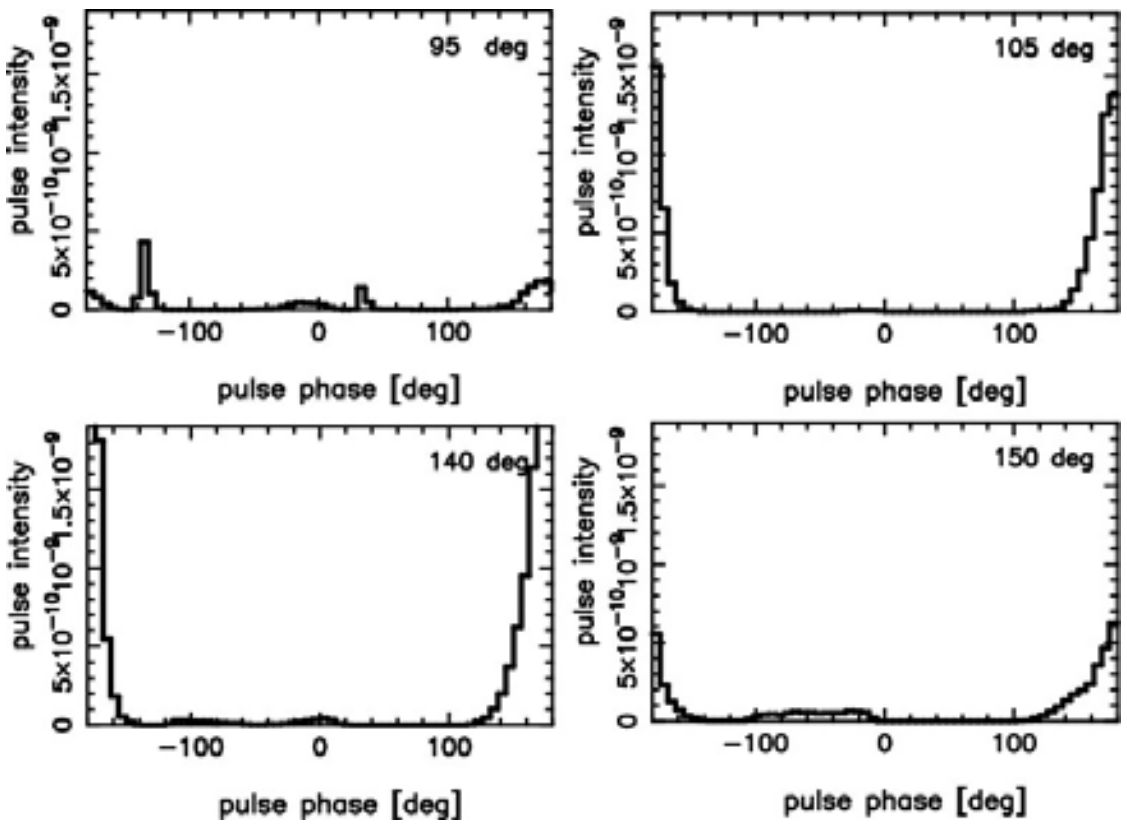

Fig. (18) Pulse profile for four different viewing angles, $\zeta_{obs}$, in the higher-altitude SG model. Note that the typical ordinate values ($\sim 10^{-10}$) are $10^3$ times less than those in figure 8 ($\sim 10^{-7}$).

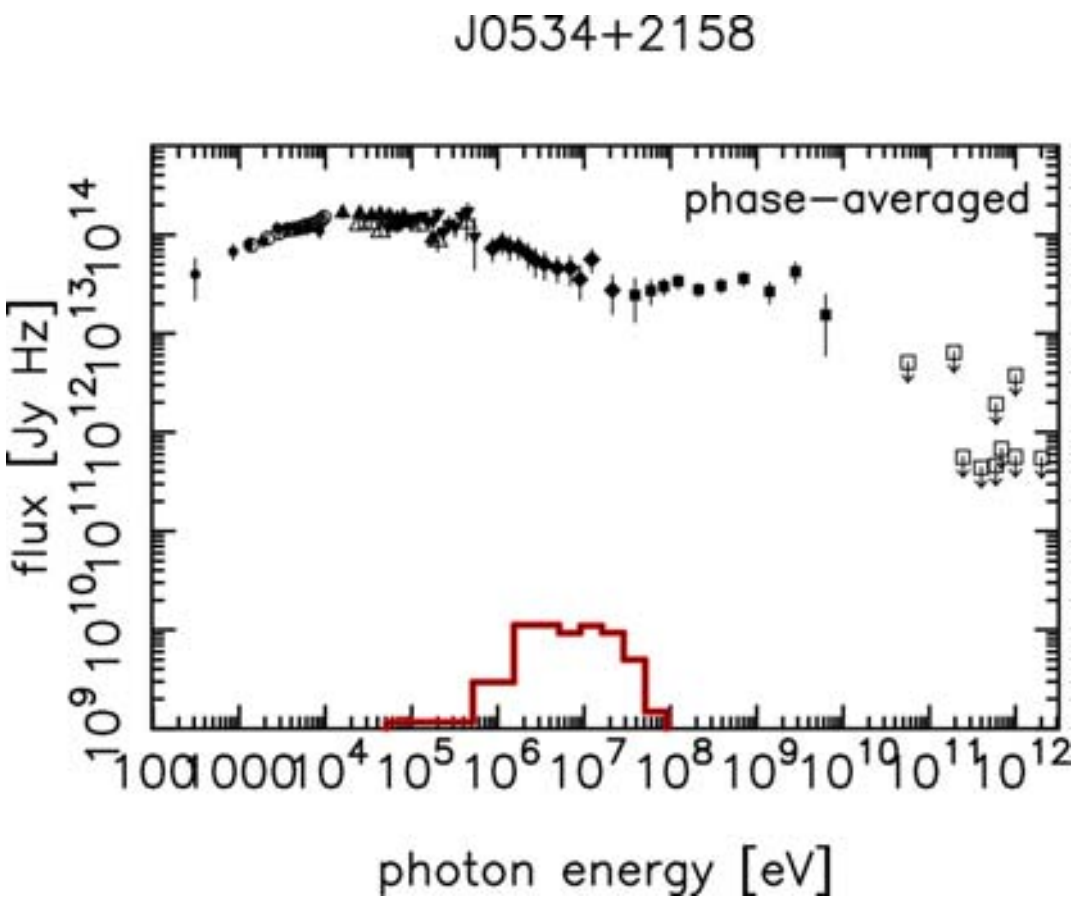

Fig. (19) Higher-altitude SG prediction of the phase-averaged spectrum of the Crab pulsar for $\zeta_{obs}$=60$^o$. Flux is negligible above 100 MeV. Lower-altitude SG emission, which is due to the synchrotron process from the cascaded pairs above the polar-cap PFF and becomes important below 10 MeV, is not depicted to clarify the higher-altitude SG component.

## 5 Discussion

To sum up, we have examined the electrodynamic structure of particle accelerators in a 3-D pulsar magnetosphere, by solving the set of Maxwell and Boltzmann equations. The accelerator (or the gap) can be solved, if we specify the following four parameters: pulsar period, period derivative (or the magnetic dipole moment), magnetic inclination angle, and the NS surface temperature, without introducing any artificial assumptions. The obtained solution (coincidentally) corresponds to a quantitative extension of previous phenomenological OG models, which adopt the acceleration electric field solved in a vacuum gap while assume non-vacuum plasma density to compute emissions. If we additionally specify the distance and the observer's viewing angle, we can predict the pulse profiles and the phase-resolved spectra of arbitrary pulsars. We apply the scheme to the Crab pulsar and obtain consistent results with observations (at least qualitatively). We also apply the same numerical scheme to the SG model, introducing an additional assumption on the real charge density and find that the SG model can predict at most 0.1 % of the observed gamma-ray flux, because its trans-field thickness is severely limited due to the pair-free condition, which is necessary not to violate the space-charge-limited flow assumption in the lower altitudes.

### *5.1. Similarities between vacuum OG model and pair-free SG model*

It is demonstrated in section 3.2 in H06 that a 2-D OG has the produced electric current that is less than the Goldreich-Julian value if $h_m$<0.047 and that the predicted $\gamma$-ray luminosity is less than 1% of the super-GJ solutions (fig. 8 in H06). This conclusion is unchanged in the present 3-D analysis, except that $h_m$ varies as a function of $s$ and $\varphi_*$.

The situation is similar for the SG model. That is, if we impose a pair-free condition, we have to adopt $h_m$<0.02 in most portion of the SG. The allowed $h_m$ becomes less than half of the limit (0.047) for an OG to be nearly vacuum, because the SG has electrons extracted from the NS surface at the GJ rate. These outgoing electrons increase the $\gamma$-$\gamma$ pair production between the outwardly propagating curvature photons and the surface thermal X-rays compared with the nearly vacuum (i.e., sub-GJ) OG case, which has a much smaller outward particle flux. Since the produced positrons return to emit copious curvature $\gamma$-rays inwards, the increased pair production in the higher-altitude SG significantly suppresses $h_m$ compared to a nearly vacuum OG. Thus, the predicted SG $\gamma$-ray flux becomes even smaller than the nearly vacuum OG case (left panel of fig. 8 in H06).

In short, neither the vacuum OG model nor the pair-free SG model can explain the observed $\gamma$-ray flux. We must consider copious pair production in the gap, which allows much greater $h_m$. In this paper, we naturally obtain this kind of solution from the set of Maxwell and Boltzmann equations under appropriate boundary conditions.

### *5.2. Assumed real charge density in SG model*

Let us briefly discuss how the results change if we adopt another real charge density in the higher-altitude SG model, instead of equation (27). If we adopt a larger (or a smaller) $|\rho_{eff}|$ than equation (27), the enhanced (or reduced) $E_{||}$ results in a smaller (or larger) $h_m$ such that the SG is kept pair-free. As a result, it turns out that the potential drop, and hence the photon flux, little depends on the assumed strength of $|\rho_{eff}|$. This kind of negative feedback effect has been, in fact, well-known in OG theories (section 4.2 in H06). On these grounds, we may consider that the OG and the SG (and even the PC) models are essentially the same, being derived from the same set of basic equations with slightly different boundary conditions (and additional assumptions in the case of the SG model). Only human may be trying to discriminate them.

*5.3. Source of pulsar winds*

In the solution that quantifies the OG model (section 3), the pair production rate in the entire magnetosphere is found to be ~$5\times10^{38}$ $s^{-1}$. This value is less than the constraints that arise from consideration of magnetic dissipation in the wind zone ~$10^{40}$ $s^{-1}$ (Kirk & Skjaeraasen 2003) and of Crab nebula's radio synchrotron emission $3\times10^{40}$ $s^{-1}$ (Arons 2004). It is likely that any OG model cannot explain such a large production rate of electron-positron pairs with outward initial momenta. (If they are produced inwardly with typical Lorentz factor of ~$10^3$, they cannot return outwards, by e.g., radiation pressure from the NS surface.) Thus, it is possible that most of the wind particles are produced (in the lower altitudes) above the PFF of a PC accelerator, which is located near the magnetic pole and does not exist on the same field lines as the OG resides. It is still an unsettled question how the PC and OG accelerators co-exist in the magnetosphere with the electric current determined by global conditions (e.g., Mestel & Shibata 1994; Spitkovsky 2006).

*5.4. Application to other pulsars*

In the present paper, we have concentrated on the electrodynamics of the two representative pulsar high-energy emission theories, OG and SG models, focusing on the application to the Crab pulsar, because we consider that theoretical consistency is more important than comparison with observations. Nevertheless, since we have developed the scheme as general as possible, the same numerical code can be applied to other rotation-powered pulsars without any modifications. In subsequent papers, we will present more results on other young pulsars, as well as on middle-aged, millisecond, and possible radio pulsars, to explain the data obtained by e.g., ROSAT, Chandra, XMM, ASCA, BeppoSAX, CGRO, AGILE and to predict potential pulsars for GLAST and MAGIC observations.

## ACKNOWLEDGEMENTS

Special thanks are due to Dr. A. K. Harding for many valuable comments and criticisms. The author is also indebted to Drs. K. S. Cheng, H. K. Chang, J. Takata for fruitful discussion. This work is supported by the Theoretical Institute for Advance Research in Astrophysics (TIARA) operated under Academia Sinica and the National Science Council Excellence Projects program in Taiwan administered through grant number NSC 95-2752-M-007-006-PAE.